%**start of header
\gdef\islinuxolivetti{F}
%%%%%%%%%%%%%%%%%%%%%%%%%
%%%%layout %%%%%%%%%%%%%%
%%%%%%%%%%%%%%%%%%%%%%%%%
\magnification\magstep1
%for dvips with figures maybe uncomment next line

%%%%%%%%%%%%%%%%%%%%%%%
%%%% new parameters %%%
%%%%%%%%%%%%%%%%%%%%%%%
\newdimen\papwidth
\newdimen\papheight
\newskip\beforesectionskipamount  %how much to skip before section title
\newskip\sectionskipamount %how much to skip after section title
\def\sectionskip{\vskip\sectionskipamount}
\def\beforesectionskip{\vskip\beforesectionskipamount}
%%%%%%%%%%%%%%%%%%%%%%%
%%%% paper %%%%%%%%%%%%
%%%%%%%%%%%%%%%%%%%%%%%
\papwidth=16truecm
\papheight=22truecm
\voffset=0.4truecm
\hoffset=0.4truecm
%%%%%%%install variables%%%%%%%%%%%%%%%%%%%
\hsize=\papwidth
\vsize=\papheight
%%%%%%%%%%%%%%%%%%%%%%%%%
%%%%%% headline %%%%%%%%%
%%%%%%%%%%%%%%%%%%%%%%%%%
\nopagenumbers
\headline={\ifnum\pageno>1 {\hss\tenrm-\ \folio\ -\hss} \else
{\hfill}\fi}
%%%%%%%%%%%%%%%%%%%%%%%%%
\newdimen\texpscorrection
\texpscorrection=0.15truecm %must be 0.15truecm in ps_fonts
%%%%%%%%%%%%%%%%%%%%%%%%
%%%%%%% fontsizes %%%%%%
%%%%%%%%%%%%%%%%%%%%%%%%

\def\sectionsize{\twelvepoint}
\def\sectiontype{\bf}
\def\subsectionsize{}
\def\subsectiontype{\bf}
\def\em{\sl}%will be italic in reality
%%%%%%%%%%%%%%%%%%%%%
\newfam\truecmsy
\newfam\truecmr
\newfam\msbfam
\newfam\scriptfam
\newfam\truecmsy
%%%%%%%%%%%%%%%%%%%%%%%%%%%%%%%%%%%%%
\newskip\ttglue 
%%%%%%%%%%%%%%%%%%%%%%%%%%%%%%%%%%%%%%%%%%%
% Font for LINUX
%%%%%%%%%%%%%%%%%%%%%%%%%%%%%%%%%%%%%%%%%%%%%
\if T\islinuxolivetti
\papheight=11truecm

% Times-Roman
\font\twelverm=cmr12
\font\tenrm=cmr10
%\font\ninerm=cmr9
\font\eightrm=cmr8
\font\sevenrm=cmr7
\font\sixrm=cmr6
\font\fiverm=cmr5

% Times-Bold
\font\twelvebf=cmbx12
\font\tenbf=cmbx10
%\font\ninebf=cmbx9
\font\eightbf=cmbx8
\font\sevenbf=cmbx7
\font\sixbf=cmbx6
\font\fivebf=cmbx5

% Times-Italic
\font\twelveit=cmti12
\font\tenit=cmti10
%\font\nineit=cmti9
\font\eightit=cmti8
\font\sevenit=cmti7
\font\sixit=cmti6
\font\fiveit=cmti5

% Times-Oblique(slanted)
\font\twelvesl=cmsl12
\font\tensl=cmsl10
%\font\ninesl=cmsl9
\font\eightsl=cmsl8
\font\sevensl=cmsl7
\font\sixsl=cmsl6
\font\fivesl=cmsl5

% Math-Italic
\font\twelvei=cmmi12
\font\teni=cmmi10
%\font\ninei=cmmi9
\font\eighti=cmmi8
\font\seveni=cmmi7
\font\sixi=cmmi6
\font\fivei=cmmi5

% Math-Symbols
\font\twelvesy=cmsy10	at	12pt
\font\tensy=cmsy10
%\font\ninesy=cmsy9
\font\eightsy=cmsy8
\font\sevensy=cmsy7
\font\sixsy=cmsy6
\font\fivesy=cmsy5
\font\twelvetruecmsy=cmsy10	at	12pt
\font\tentruecmsy=cmsy10
%\font\ninetruecmsy=cmsy9
\font\eighttruecmsy=cmsy8
\font\seventruecmsy=cmsy7
\font\sixtruecmsy=cmsy6
\font\fivetruecmsy=cmsy5

% CM-Roman
\font\twelvetruecmr=cmr12
\font\tentruecmr=cmr10
%\font\ninetruecmr=cmr9
\font\eighttruecmr=cmr8
\font\seventruecmr=cmr7
\font\sixtruecmr=cmr6
\font\fivetruecmr=cmr5

% Math-Boldfaces
\font\twelvebf=cmbx12
\font\tenbf=cmbx10
%\font\ninebf=cmbx9
\font\eightbf=cmbx8
\font\sevenbf=cmbx7
\font\sixbf=cmbx6
\font\fivebf=cmbx5

% Teletype
\font\twelvett=cmtt12
\font\tentt=cmtt10
%\font\ninett=cmtt9
\font\eighttt=cmtt8

% Big Math Symbols
\font\twelveex=cmex10	at	12pt
\font\tenex=cmex10
%\font\nineex=cmex9

% AMS Math Symbols
\font\twelvemsb=msbm10	at	12pt
\font\tenmsb=msbm10
%\font\ninemsb=msbm9
\font\eightmsb=msbm8
\font\sevenmsb=msbm7
\font\sixmsb=msbm6
\font\fivemsb=msbm5

%%% Fraktur
%%\newfam\frakfam
%%\font\twelvefrm=eufm10	at	12pt
%%\font\tenfrm=eufm10
%%%\font\ninefrm=eufm9
%%\font\eightfrm=eufm8
%%\font\sevenfrm=eufm7
%%\font\sixfrm=eufm6
%%\font\fivefrm=eufm5
%%
%%% Bold Fraktur
%%\newfam\frakbfam
%%\font\twelvefrb=eufb10 at 12pt
%%\font\tenfrb=eufb10
%%%\font\ninefrb=eufb9
%%\font\eightfrb=eufb8
%%\font\sevenfrb=eufb7
%%\font\sixfrb=eufb6
%%\font\fivefrb=eufb5
%%
% Script-Faces
\font\twelvescr=eusm10 at 12pt
\font\tenscr=eusm10
%\font\ninescr=eusm9
\font\eightscr=eusm8
\font\sevenscr=eusm7
\font\sixscr=eusm6
\font\fivescr=eusm5
\fi
\if F\islinuxolivetti
%%%%%%%%%%%%%%%%%%%%%%%%%%%%%%%%%%%%%%
% Font mapping for postscript fonts.
%%%%%%%%%%%%%%%%%%%%%%%%%%%%%%%%%%%%%%
% Times-Roman
\font\twelverm=ptmr	at	12pt
\font\tenrm=ptmr	at	10pt
%\font\ninerm=ptmr	at	9pt
\font\eightrm=ptmr	at	8pt
\font\sevenrm=ptmr	at	7pt
\font\sixrm=ptmr	at	6pt
\font\fiverm=ptmr	at	5pt

% Times-Bold
\font\twelvebf=ptmb	at	12pt
\font\tenbf=ptmb	at	10pt
%\font\ninebf=ptmb	at	9pt
\font\eightbf=ptmb	at	8pt
\font\sevenbf=ptmb	at	7pt
\font\sixbf=ptmb	at	6pt
\font\fivebf=ptmb	at	5pt

% Times-Italic
\font\twelveit=ptmri	at	12pt
\font\tenit=ptmri	at	10pt
%\font\nineit=ptmri	at	9pt
\font\eightit=ptmri	at	8pt
\font\sevenit=ptmri	at	7pt
\font\sixit=ptmri	at	6pt
\font\fiveit=ptmri	at	5pt

% Times-Oblique(slanted)
\font\twelvesl=ptmro	at	12pt
\font\tensl=ptmro	at	10pt
%\font\ninesl=ptmro	at	9pt
\font\eightsl=ptmro	at	8pt
\font\sevensl=ptmro	at	7pt
\font\sixsl=ptmro	at	6pt
\font\fivesl=ptmro	at	5pt

% Math-Italic
\font\twelvei=cmmi12
\font\teni=cmmi10
%\font\ninei=cmmi9
\font\eighti=cmmi8
\font\seveni=cmmi7
\font\sixi=cmmi6
\font\fivei=cmmi5

% Math-Symbols
\font\twelvesy=cmsy10	at	12pt
\font\tensy=cmsy10
%\font\ninesy=cmsy9
\font\eightsy=cmsy8
\font\sevensy=cmsy7
\font\sixsy=cmsy6
\font\fivesy=cmsy5
\font\twelvetruecmsy=cmsy10	at	12pt
\font\tentruecmsy=cmsy10
%\font\ninetruecmsy=cmsy9
\font\eighttruecmsy=cmsy8
\font\seventruecmsy=cmsy7
\font\sixtruecmsy=cmsy6
\font\fivetruecmsy=cmsy5

% CM-Roman
\font\twelvetruecmr=cmr12
\font\tentruecmr=cmr10
%\font\ninetruecmr=cmr9
\font\eighttruecmr=cmr8
\font\seventruecmr=cmr7
\font\sixtruecmr=cmr6
\font\fivetruecmr=cmr5

% Math-Boldfaces
\font\twelvebf=cmbx12
\font\tenbf=cmbx10
%\font\ninebf=cmbx9
\font\eightbf=cmbx8
\font\sevenbf=cmbx7
\font\sixbf=cmbx6
\font\fivebf=cmbx5

% Teletype
\font\twelvett=cmtt12
\font\tentt=cmtt10
%\font\ninett=cmtt9
\font\eighttt=cmtt8

% Big Math Symbols
\font\twelveex=cmex10	at	12pt
\font\tenex=cmex10
%\font\nineex=cmex9

% AMS Math Symbols
\font\twelvemsb=msbm10	at	12pt
\font\tenmsb=msbm10
%\font\ninemsb=msbm9
\font\eightmsb=msbm8
\font\sevenmsb=msbm7
\font\sixmsb=msbm6
\font\fivemsb=msbm5

%%% Fraktur
%%\newfam\frakfam
%%\font\twelvefrm=eufm10	at	12pt
%%\font\tenfrm=eufm10
%%%\font\ninefrm=eufm9
%%\font\eightfrm=eufm8
%%\font\sevenfrm=eufm7
%%\font\sixfrm=eufm6
%%\font\fivefrm=eufm5
%%
%%% Bold Fraktur
%%\newfam\frakbfam
%%\font\twelvefrb=eufb10 at 12pt
%%\font\tenfrb=eufb10
%%%\font\ninefrb=eufb9
%%\font\eightfrb=eufb8
%%\font\sevenfrb=eufb7
%%\font\sixfrb=eufb6
%%\font\fivefrb=eufb5
%%
% Script-Faces
\font\twelvescr=eusm10 at 12pt
\font\tenscr=eusm10
%\font\ninescr=eusm9
\font\eightscr=eusm8
\font\sevenscr=eusm7
\font\sixscr=eusm6
\font\fivescr=eusm5
\fi
%%%%%%%%%%%%%%%%%%%%%%%%%
%%%%preloaded fonts%%%%%%
%%%%%%%%%%%%%%%%%%%%%%%%%
\def\eightpoint{\def\rm{\fam0\eightrm}%
\textfont0=\eightrm
  \scriptfont0=\sixrm
  \scriptscriptfont0=\fiverm 
\textfont1=\eighti
  \scriptfont1=\sixi
  \scriptscriptfont1=\fivei 
\textfont2=\eightsy
  \scriptfont2=\sixsy
  \scriptscriptfont2=\fivesy 
\textfont3=\tenex
  \scriptfont3=\tenex
  \scriptscriptfont3=\tenex 
\textfont\itfam=\eightit
  \scriptfont\itfam=\sixit
  \scriptscriptfont\itfam=\fiveit 
  \def\it{\fam\itfam\eightit}%
\textfont\slfam=\eightsl
  \scriptfont\slfam=\sixsl
  \scriptscriptfont\slfam=\fivesl 
  \def\sl{\fam\slfam\eightsl}%
\textfont\ttfam=\eighttt
  \def\tt{\fam\ttfam\eighttt}%
\textfont\bffam=\eightbf
  \scriptfont\bffam=\sixbf
  \scriptscriptfont\bffam=\fivebf
  \def\bf{\fam\bffam\eightbf}%
%%\textfont\frakfam=\eightfrm
%%  \scriptfont\frakfam=\sixfrm
%%  \scriptscriptfont\frakfam=\fivefrm
%%  \def\frak{\fam\frakfam\eightfrm}%
%%\textfont\frakbfam=\eightfrb
%%  \scriptfont\frakbfam=\sixfrb
%%  \scriptscriptfont\frakbfam=\fivefrb
%%  \def\bfrak{\fam\frakbfam\eightfrb}%
\textfont\scriptfam=\eightscr
  \scriptfont\scriptfam=\sixscr
  \scriptscriptfont\scriptfam=\fivescr
  \def\script{\fam\scriptfam\eightscr}%
\textfont\msbfam=\eightmsb
  \scriptfont\msbfam=\sixmsb
  \scriptscriptfont\msbfam=\fivemsb
  \def\bb{\fam\msbfam\eightmsb}%
\textfont\truecmr=\eighttruecmr
  \scriptfont\truecmr=\sixtruecmr
  \scriptscriptfont\truecmr=\fivetruecmr
  \def\truerm{\fam\truecmr\eighttruecmr}%
\textfont\truecmsy=\eighttruecmsy
  \scriptfont\truecmsy=\sixtruecmsy
  \scriptscriptfont\truecmsy=\fivetruecmsy
\tt \ttglue=.5em plus.25em minus.15em 
\normalbaselineskip=9pt
\setbox\strutbox=\hbox{\vrule height7pt depth2pt width0pt}%
\normalbaselines
\rm
}

\def\tenpoint{\def\rm{\fam0\tenrm}%
\textfont0=\tenrm
  \scriptfont0=\sevenrm
  \scriptscriptfont0=\fiverm 
\textfont1=\teni
  \scriptfont1=\seveni
  \scriptscriptfont1=\fivei 
\textfont2=\tensy
  \scriptfont2=\sevensy
  \scriptscriptfont2=\fivesy 
\textfont3=\tenex
  \scriptfont3=\tenex
  \scriptscriptfont3=\tenex 
\textfont\itfam=\tenit
  \scriptfont\itfam=\sevenit
  \scriptscriptfont\itfam=\fiveit 
  \def\it{\fam\itfam\tenit}%
\textfont\slfam=\tensl
  \scriptfont\slfam=\sevensl
  \scriptscriptfont\slfam=\fivesl 
  \def\sl{\fam\slfam\tensl}%
\textfont\ttfam=\tentt
  \def\tt{\fam\ttfam\tentt}%
\textfont\bffam=\tenbf
  \scriptfont\bffam=\sevenbf
  \scriptscriptfont\bffam=\fivebf
  \def\bf{\fam\bffam\tenbf}%
%%\textfont\frakfam=\tenfrm
%%  \scriptfont\frakfam=\sevenfrm
%%  \scriptscriptfont\frakfam=\fivefrm
%%  \def\frak{\fam\frakfam\tenfrm}%
%%\textfont\frakbfam=\tenfrb
%%  \scriptfont\frakbfam=\sevenfrb
%%  \scriptscriptfont\frakbfam=\fivefrb
%%  \def\bfrak{\fam\frakbfam\tenfrb}%
\textfont\scriptfam=\tenscr
  \scriptfont\scriptfam=\sevenscr
  \scriptscriptfont\scriptfam=\fivescr
  \def\script{\fam\scriptfam\tenscr}%
\textfont\msbfam=\tenmsb
  \scriptfont\msbfam=\sevenmsb
  \scriptscriptfont\msbfam=\fivemsb
  \def\bb{\fam\msbfam\tenmsb}%
\textfont\truecmr=\tentruecmr
  \scriptfont\truecmr=\seventruecmr
  \scriptscriptfont\truecmr=\fivetruecmr
  \def\truerm{\fam\truecmr\tentruecmr}%
\textfont\truecmsy=\tentruecmsy
  \scriptfont\truecmsy=\seventruecmsy
  \scriptscriptfont\truecmsy=\fivetruecmsy
\tt \ttglue=.5em plus.25em minus.15em 
\normalbaselineskip=12pt
\setbox\strutbox=\hbox{\vrule height8.5pt depth3.5pt width0pt}%
\normalbaselines
\rm
}

\def\twelvepoint{\def\rm{\fam0\twelverm}%
\textfont0=\twelverm
  \scriptfont0=\tenrm
  \scriptscriptfont0=\eightrm 
\textfont1=\twelvei
  \scriptfont1=\teni
  \scriptscriptfont1=\eighti 
\textfont2=\twelvesy
  \scriptfont2=\tensy
  \scriptscriptfont2=\eightsy 
\textfont3=\twelveex
  \scriptfont3=\twelveex
  \scriptscriptfont3=\twelveex 
\textfont\itfam=\twelveit
  \scriptfont\itfam=\tenit
  \scriptscriptfont\itfam=\eightit 
  \def\it{\fam\itfam\twelveit}%
\textfont\slfam=\twelvesl
  \scriptfont\slfam=\tensl
  \scriptscriptfont\slfam=\eightsl 
  \def\sl{\fam\slfam\twelvesl}%
\textfont\ttfam=\twelvett
  \def\tt{\fam\ttfam\twelvett}%
\textfont\bffam=\twelvebf
  \scriptfont\bffam=\tenbf
  \scriptscriptfont\bffam=\eightbf
  \def\bf{\fam\bffam\twelvebf}%
%%\textfont\frakfam=\twelvefrm
%%  \scriptfont\frakfam=\tenfrm
%%  \scriptscriptfont\frakfam=\eightfrm
%%  \def\frak{\fam\frakfam\twelvefrm}%
%%\textfont\frakbfam=\twelvefrb
%%  \scriptfont\frakbfam=\tenfrb
%%  \scriptscriptfont\frakbfam=\eightfrb
%%  \def\bfrak{\fam\frakbfam\twelvefrb}%
\textfont\scriptfam=\twelvescr
  \scriptfont\scriptfam=\tenscr
  \scriptscriptfont\scriptfam=\eightscr
  \def\script{\fam\scriptfam\twelvescr}%
\textfont\msbfam=\twelvemsb
  \scriptfont\msbfam=\tenmsb
  \scriptscriptfont\msbfam=\eightmsb
  \def\bb{\fam\msbfam\twelvemsb}%
\textfont\truecmr=\twelvetruecmr
  \scriptfont\truecmr=\tentruecmr
  \scriptscriptfont\truecmr=\eighttruecmr
  \def\truerm{\fam\truecmr\twelvetruecmr}%
\textfont\truecmsy=\twelvetruecmsy
  \scriptfont\truecmsy=\tentruecmsy
  \scriptscriptfont\truecmsy=\eighttruecmsy
\tt \ttglue=.5em plus.25em minus.15em 
\setbox\strutbox=\hbox{\vrule height7pt depth2pt width0pt}%
\normalbaselineskip=15pt
\normalbaselines
\rm
}
%
%%%%%constant subscript positions%%%%%
\fontdimen16\tensy=2.7pt
%\fontdimen13\tensy=2.7pt
\fontdimen13\tensy=4.3pt
\fontdimen17\tensy=2.7pt
\fontdimen14\tensy=4.3pt
\fontdimen18\tensy=4.3pt
\fontdimen16\eightsy=2.7pt
\fontdimen13\eightsy=4.3pt
\fontdimen17\eightsy=2.7pt
\fontdimen14\eightsy=4.3pt
\fontdimen18\eightsy=4.3pt
%
%%%%%%%%%%%%%%%%%%%%%%%%%%%%%%%%%%%%%%%%%%%%%%%%%%%%%%%%%%%%%
%%%%%%%%%%%%%% redefine some math so that it is cmr %%%%%%%%%
%%%%%%%%%%%%%%%%%%%%%%%%%%%%%%%%%%%%%%%%%%%%%%%%%%%%%%%%%%%%%
\def\hexnumber#1{\ifcase#1 0\or1\or2\or3\or4\or5\or6\or7\or8\or9\or
 A\or B\or C\or D\or E\or F\fi}
\mathcode`\=="3\hexnumber\truecmr3D
\mathchardef\not="3\hexnumber\truecmsy36
\mathcode`\+="2\hexnumber\truecmr2B
\mathcode`\(="4\hexnumber\truecmr28
\mathcode`\)="5\hexnumber\truecmr29
\mathcode`\!="5\hexnumber\truecmr21
\mathcode`\(="4\hexnumber\truecmr28
\mathcode`\)="5\hexnumber\truecmr29
%\chardef`,="0\hexnum\truecmr3B

\def\tilde{\mathaccent"0\hexnumber\truecmr7E }

\def\dot{\mathaccent"0\hexnumber\truecmr5F }
\def\Phi{\mathchar"0\hexnumber\truecmr08 }
\def\Gamma {\mathchar"0\hexnumber\truecmr00 }
\def\Delta {\mathchar"0\hexnumber\truecmr01 }
\def\Theta {\mathchar"0\hexnumber\truecmr02 }
\def\Lambda{\mathchar"0\hexnumber\truecmr03 }
\def\Xi {\mathchar"0\hexnumber\truecmr04 }
\def\Pi{\mathchar"0\hexnumber\truecmr05 }
\def\Sigma{\mathchar"0\hexnumber\truecmr06 }
\def\Upsilon {\mathchar"0\hexnumber\truecmr07 }
\def\Phi {\mathchar"0\hexnumber\truecmr08 }
\def\Psi {\mathchar"0\hexnumber\truecmr09 }
\def\Omega{\mathchar"0\hexnumber\truecmr0A }
%%%%%%%%%%%%%%%%%%%%%%%%%%%%%%%%%%%%%%%%%%%%%%
%%% macros  for cross reference %%%%%%%%%%%%%%
%%%%%%%%%%%%%%%%%%%%%%%%%%%%%%%%%%%%%%%%%%%%%%
%%
%%  counters %%%
%%  
\newcount\EQNcount \EQNcount=1
\newcount\CLAIMcount \CLAIMcount=1
\newcount\SECTIONcount \SECTIONcount=0
\newcount\SUBSECTIONcount \SUBSECTIONcount=1
%%
%% defining the symbolic value
%%
\def\ifff(#1,#2,#3){\ifundefined{#1#2}%
\expandafter\xdef\csname #1#2\endcsname{#3}\else%
\immediate\write16{!!!!!doubly defined #1,#2}\fi}
\def\NEWDEF #1,#2,#3 {\ifff({#1},{#2},{#3})}
\def\actualnumber{\number\SECTIONcount}
\def\EQ(#1){\lmargin(#1)\eqno\tag(#1)}
\def\NR(#1){&\lmargin(#1)\tag(#1)\cr}  %the same as &\tag(xx)\cr in eqalignno
\def\tag(#1){\lmargin(#1)({\rm \actualnumber}.\number\EQNcount)
 \NEWDEF e,#1,(\actualnumber.\number\EQNcount)
\global\advance\EQNcount by 1
%\immediate\write16{ EQ \equ(#1):#1  }
}
\def\SECT(#1)#2\par{\lmargin(#1)\SECTION#2\par
\NEWDEF s,#1,{\actualnumber}
}
\def\SUBSECT(#1)#2\par{\lmargin(#1)
\SUBSECTION#2\par 
\NEWDEF s,#1,{\actualnumber.\number\SUBSECTIONcount}
}
%%%% the actual macro %%%%%%
\def\CLAIM #1(#2) #3\par{
\vskip.1in\medbreak\noindent
{\lmargin(#2)\bf #1\ \actualnumber.\number\CLAIMcount.} {\sl #3}\par
\NEWDEF c,#2,{#1\ \actualnumber.\number\CLAIMcount}
\global\advance\CLAIMcount by 1
\ifdim\lastskip<\medskipamount
\removelastskip\penalty55\medskip\fi}
\def\CLAIMNONR #1(#2) #3\par{
\vskip.1in\medbreak\noindent
{\lmargin(#2)\bf #1.} {\sl #3}\par
\NEWDEF c,#2,{#1}
\global\advance\CLAIMcount by 1
\ifdim\lastskip<\medskipamount
\removelastskip\penalty55\medskip\fi}
\def\SECTION#1\par{\vskip0pt plus.3\vsize\penalty-75
    \vskip0pt plus -.3\vsize
    \global\advance\SECTIONcount by 1
    \beforesectionskip\noindent
{\sectionsize\sectiontype \actualnumber.\ #1}
    \EQNcount=1
    \CLAIMcount=1
    \SUBSECTIONcount=1
    \nobreak\sectionskip\noindent}
\def\SECTIONNONR#1\par{\vskip0pt plus.3\vsize\penalty-75
    \vskip0pt plus -.3\vsize
    \global\advance\SECTIONcount by 1
    \beforesectionskip\noindent
{\sectionsize\sectiontype  #1}
     \EQNcount=1
     \CLAIMcount=1
     \SUBSECTIONcount=1
     \nobreak\sectionskip\noindent}
\def\SUBSECTION#1\par{\vskip0pt plus.2\vsize\penalty-75%
    \vskip0pt plus -.2\vsize%
    \beforesectionskip\noindent%
{\subsectionsize\subsectiontype \actualnumber.\number\SUBSECTIONcount.\ #1}
    \global\advance\SUBSECTIONcount by 1
    \nobreak\sectionskip\noindent}
\def\SUBSECTIONNONR#1\par{\vskip0pt plus.2\vsize\penalty-75
    \vskip0pt plus -.2\vsize
\beforesectionskip\noindent
{\subsectionsize\subsectiontype #1}
    \nobreak\sectionskip\noindent\noindent}
%%
%%  referring to something
%%
\def\ifundefined#1{\expandafter\ifx\csname#1\endcsname\relax}
\def\equ(#1){\ifundefined{e#1}$\spadesuit$#1\else\csname e#1\endcsname\fi}
\def\clm(#1){\ifundefined{c#1}$\spadesuit$#1\else\csname c#1\endcsname\fi}
\def\sec(#1){\ifundefined{s#1}$\spadesuit$#1
\else Section \csname s#1\endcsname\fi}
%%%%%%%%%%%%%TITLE PAGE%%%%%%%%%%%%%%%%%%%%
\let\endarg=\par
\def\finish{\def\endarg{\par\endgroup}}
\def\start{\endarg\begingroup}

 \def\beginFROM{\start\parskip=0pt\vskip\baselineskip
\def\finish{\def\endarg{\egroup\par\endgroup}}
  \vbox\bgroup\obeylines\eightpoint\em\finish}

\def\ABSTRACT#1\par{
\vskip 1in {\noindent\sectionsize\sectiontype Abstract.} #1 \par}

%%%%%%%%%%% The today mechanism %%%%%%%%%%%%%%%%%
\def\TODAY{\number\day~\ifcase\month\or January \or February \or March \or
April \or May \or June
\or July \or August \or September \or October \or November \or December \fi
\number\year\timecount=\number\time
\divide\timecount by 60
}
\newcount\timecount
\def\DRAFT{\def\lmargin(##1){\strut\vadjust{\kern-\strutdepth
\vtop to \strutdepth{
\baselineskip\strutdepth\vss\rlap{\kern-1.2 truecm\eightpoint{##1}}}}}
\font\footfont=cmti7
\footline={{\footfont \hfil File:\jobname, \TODAY,  \number\timecount h}}
}
%%%subitem an item in a vbox%%%%
\newbox\strutboxJPE
\setbox\strutboxJPE=\hbox{\strut}
\def\subitem#1#2\par{\vskip\baselineskip\vskip-\ht\strutboxJPE{\item{#1}#2}}
\gdef\strutdepth{\dp\strutbox}
\def\lmargin(#1){}
%%%%%%%%%%%%%%%%BIBLIOGRAPHY%%%%%%%%%%%%%%%%%%%%
%%%%%%%%%%%%%%%%%%%%%%%%%%%%%%%%%%%%%%%%%%%%%%%%
\def\period{\unskip.\spacefactor3000 { }}
%
% ...invisible stuff
%
\newbox\noboxJPE
\newbox\byboxJPE
\newbox\paperboxJPE
\newbox\yrboxJPE
\newbox\jourboxJPE
\newbox\pagesboxJPE
\newbox\volboxJPE
\newbox\preprintboxJPE
\newbox\toappearboxJPE
\newbox\bookboxJPE
\newbox\bybookboxJPE
\newbox\publisherboxJPE
\newbox\inprintboxJPE
\def\refclearJPE{
   \setbox\noboxJPE=\null             \gdef\isnoJPE{F}
   \setbox\byboxJPE=\null             \gdef\isbyJPE{F}
   \setbox\paperboxJPE=\null          \gdef\ispaperJPE{F}
   \setbox\yrboxJPE=\null             \gdef\isyrJPE{F}
   \setbox\jourboxJPE=\null           \gdef\isjourJPE{F}
   \setbox\pagesboxJPE=\null          \gdef\ispagesJPE{F}
   \setbox\volboxJPE=\null            \gdef\isvolJPE{F}
   \setbox\preprintboxJPE=\null       \gdef\ispreprintJPE{F}
   \setbox\toappearboxJPE=\null       \gdef\istoappearJPE{F}
   \setbox\inprintboxJPE=\null        \gdef\isinprintJPE{F}
   \setbox\bookboxJPE=\null           \gdef\isbookJPE{F}  \gdef\isinbookJPE{F}
     
   \setbox\bybookboxJPE=\null         \gdef\isbybookJPE{F}
   \setbox\publisherboxJPE=\null      \gdef\ispublisherJPE{F}
     
}
\def\widestlabel#1{\setbox0=\hbox{#1\enspace}\refindent=\wd0\relax}
\def\ref{\refclearJPE\bgroup}
\def\no   {\egroup\gdef\isnoJPE{T}\setbox\noboxJPE=\hbox\bgroup}
\def\by   {\egroup\gdef\isbyJPE{T}\setbox\byboxJPE=\hbox\bgroup}
\def\paper{\egroup\gdef\ispaperJPE{T}\setbox\paperboxJPE=\hbox\bgroup}
\def\yr{\egroup\gdef\isyrJPE{T}\setbox\yrboxJPE=\hbox\bgroup}
\def\jour{\egroup\gdef\isjourJPE{T}\setbox\jourboxJPE=\hbox\bgroup}
\def\pages{\egroup\gdef\ispagesJPE{T}\setbox\pagesboxJPE=\hbox\bgroup}
\def\vol{\egroup\gdef\isvolJPE{T}\setbox\volboxJPE=\hbox\bgroup\bf}
\def\preprint{\egroup\gdef
\ispreprintJPE{T}\setbox\preprintboxJPE=\hbox\bgroup}
\def\toappear{\egroup\gdef
\istoappearJPE{T}\setbox\toappearboxJPE=\hbox\bgroup}
\def\inprint{\egroup\gdef
\isinprintJPE{T}\setbox\inprintboxJPE=\hbox\bgroup}
\def\book{\egroup\gdef\isbookJPE{T}\setbox\bookboxJPE=\hbox\bgroup\em}
\def\publisher{\egroup\gdef
\ispublisherJPE{T}\setbox\publisherboxJPE=\hbox\bgroup}
\def\inbook{\egroup\gdef\isinbookJPE{T}\setbox\bookboxJPE=\hbox\bgroup\em}
\def\bybook{\egroup\gdef\isbybookJPE{T}\setbox\bybookboxJPE=\hbox\bgroup}
\newdimen\refindent
\refindent=5em
\def\endref{\egroup \sfcode`.=1000
 \if T\isnoJPE
% \setbox0=\hbox{[\unhbox\noboxJPE\unskip]\hss\unskip\enspace}%
%   \ifdim\refindent<\wd0\relax
%      \message{\string\refno: reference is wider than
%               you pretended when using \string\widestlabel.}%
%   \fi
 \hangindent\refindent\hangafter=1
      \noindent\hbox to\refindent{[\unhbox\noboxJPE\unskip]\hss}\ignorespaces
     \else  \noindent    \fi
% \if T\isnoJPE  \item{[\unhbox\noboxJPE\unskip]}
%     \else  \noindent    \fi
 \if T\isbyJPE    \unhbox\byboxJPE\unskip: \fi
 \if T\ispaperJPE \unhbox\paperboxJPE\unskip\period \fi
 \if T\isbookJPE {\it\unhbox\bookboxJPE\unskip}\if T\ispublisherJPE, \else.
\fi\fi
 \if T\isinbookJPE In {\it\unhbox\bookboxJPE\unskip}\if T\isbybookJPE,
\else\period \fi\fi
 \if T\isbybookJPE  (\unhbox\bybookboxJPE\unskip)\period \fi
 \if T\ispublisherJPE \unhbox\publisherboxJPE\unskip \if T\isjourJPE, \else\if
T\isyrJPE \  \else\period \fi\fi\fi
 \if T\istoappearJPE (To appear)\period \fi
 \if T\ispreprintJPE Pre\-print\period \fi
 \if T\isjourJPE    \unhbox\jourboxJPE\unskip\ \fi
 \if T\isvolJPE     \unhbox\volboxJPE\unskip\if T\ispagesJPE, \else\ \fi\fi
 \if T\ispagesJPE   \unhbox\pagesboxJPE\unskip\  \fi
 \if T\isyrJPE      (\unhbox\yrboxJPE\unskip)\period \fi
 \if T\isinprintJPE (in print)\period \fi
\filbreak
}
\def\hexnumber#1{\ifcase#1 0\or1\or2\or3\or4\or5\or6\or7\or8\or9\or
 A\or B\or C\or D\or E\or F\fi}
\textfont\msbfam=\tenmsb
\scriptfont\msbfam=\sevenmsb
\scriptscriptfont\msbfam=\fivemsb
\mathchardef\varkappa="0\hexnumber\msbfam7B
%%%%%%%%%%%%%%%%%%%%%%%%%%%%%%%%%%%%%%%%%%%%%%%%%%%%%%%%
%%%%%%%%%%  Figures %%%%%%%%%%%%%%%%%%%%%%%%%%%%%%%%%%%%
%%%%%%%%%%%%%%%%%%%%%%%%%%%%%%%%%%%%%%%%%%%%%%%%%%%%%%%%
\newcount\FIGUREcount \FIGUREcount=0
\newdimen\figcenter
%%%%%%%%%%%%%%%%%%%%%%%%%%%%%%%%%%%%%%%%%%%%%%%%%%%%%%%%
\def\fig(#1){\ifundefined{fig#1}%
\global\advance\FIGUREcount by 1%
\NEWDEF fig,#1,{Fig.\ \number\FIGUREcount}
\immediate\write16{ FIG \number\FIGUREcount : #1}
\fi
\csname fig#1\endcsname\relax}
%%%%%%%%%%%%%%%%%%%%%%%%%%%%%%%%%%%%%%%%%%%
%figure 1=psfile=NAME 2=height (in cm) 3=width (in cm) 4=caption  
%%%%%%%%%%%%%%%%%%%%%%%%%%%%%%%%%%%%%%%%%%%%%%%%%%%%%%%
\def\figure #1 #2 #3 #4\cr{\null%
\ifundefined{fig#1}%
\global\advance\FIGUREcount by 1%
\NEWDEF fig,#1,{Fig.\ \number\FIGUREcount}
\immediate\write16{  FIG \number\FIGUREcount : #1}
\fi
{\goodbreak\figcenter=\hsize\relax
\advance\figcenter by -#3truecm
\divide\figcenter by 2
\midinsert\vskip #2truecm\noindent\hskip\figcenter
\includegraphics{#1}\vskip 0.8truecm\noindent \vbox{\eightpoint\noindent
{\bf\fig(#1)}: #4}\endinsert}}
%%%%%%%%%%%%%%%%%%%%%%%%%%%%%%%%%%%%%%%%%%%%%%%%%%%%%%%%%
%figurewithtex 1=psfile=NAME 2=texfile 3=height (in cm) 4=width
%(in cm) 5=caption
%%%%%%%%%%%%%%%%%%%%%%%%%%%%%%%%%%%%%%%%%%%%%%%%%%%%%%%%%
\def\figurewithtex #1 #2 #3 #4 #5\cr{\null%
\ifundefined{fig#1}%
\global\advance\FIGUREcount by 1%
\NEWDEF fig,#1,{Fig.\ \csname fig#1\endcsname}
\immediate\write16{ FIG \number\FIGUREcount: #1}
\fi
{\goodbreak\figcenter=\hsize\relax
\advance\figcenter by -#4truecm
\divide\figcenter by 2
\midinsert\vskip #3truecm\noindent\hskip\figcenter
\includegraphics{#1}{\hskip\texpscorrection\input #2 }\vskip 0.8truecm\noindent \vbox{\eightpoint\noindent
{\bf\fig(#1)}: #5}\endinsert}}
%%%%%%%%%%%%%%%%%%%%%%%%%%%%%%%%%%%%%%%%%%%%%%%%%%%%%%%%%%%%%%%%%%
%figurewithtexplus 1=psfile=NAME 2=texfile 3=height (in cm) 4=width
%(in cm) 5=dist figure-caption 6=caption
%%%%%%%%%%%%%%%%%%%%%%%%%%%%%%%%%%%%%%%%%%%%%%%%%%%%%%%%%%%%%%%%%%
\def\figurewithtexplus #1 #2 #3 #4 #5 #6\cr{\null%
\ifundefined{fig#1}%
\global\advance\FIGUREcount by 1%
\NEWDEF fig,#1,{Fig.\ \number\FIGUREcount}
\immediate\write16{ FIG \number\FIGUREcount: #1}
\fi
{\goodbreak\figcenter=\hsize\relax
\advance\figcenter by -#4truecm
%\advance\figcenter by -#4truecm
\divide\figcenter by 2
\midinsert\vskip #3truecm\noindent\hskip\figcenter
\includegraphics{#1}{\hskip\texpscorrection\input #2 }\vskip #5truecm\noindent \vbox{\eightpoint\noindent
{\bf\fig(#1)}: #6}\endinsert}}
%%%%%%%%%%%%%%%%%%%%%%%%%%%%%%%%%%%%%%%%%%%%%%%%%%%%%%%
\catcode`@=11
\def\footnote#1{\let\@sf\empty % parameter #2 (the text) is read later
  \ifhmode\edef\@sf{\spacefactor\the\spacefactor}\/\fi
  #1\@sf\vfootnote{#1}}
\def\vfootnote#1{\insert\footins\bgroup\eightpoint
  \interlinepenalty\interfootnotelinepenalty
  \splittopskip\ht\strutbox % top baseline for broken footnotes
  \splitmaxdepth\dp\strutbox \floatingpenalty\@MM
  \leftskip\z@skip \rightskip\z@skip \spaceskip\z@skip \xspaceskip\z@skip
  \textindent{#1}\footstrut\futurelet\next\fo@t}
\def\fo@t{\ifcat\bgroup\noexpand\next \let\next\f@@t
  \else\let\next\f@t\fi \next}
\def\f@@t{\bgroup\aftergroup\@foot\let\next}
\def\f@t#1{#1\@foot}
\def\@foot{\strut\egroup}
\def\footstrut{\vbox to\splittopskip{}}
\skip\footins=\bigskipamount % space added when footnote is present
\count\footins=1000 % footnote magnification factor (1 to 1)
\dimen\footins=8in % maximum footnotes per page
\catcode`@=12 % at signs are no longer letters
%%%%%%%%%%%%%%%%%%%%%%%
%%%  math symbols %%%%%
%%%%%%%%%%%%%%%%%%%%%%%
\def\HB {\hfill\break}
\def\AA{{\script A}}

\def\CC{{\script C}}
\def\EE{{\script E}}
\def\HH{{\script H}}
\def\LL{{\script L}}
\def\MM{{\script M}}

\def\OO{{\script O}}
\def\PP{{\script P}}

\def\HALF{{\textstyle{1\over 2}}}
%%%%%%%%%%%%%%%%%other%%%%%%%%%%%%%%%%%%%%

\def\QED{\hfill\smallskip
         \line{$\hfill{\vcenter{\vbox{\hrule height 0.2pt
	\hbox{\vrule width 0.2pt height 1.8ex \kern 1.8ex
		\vrule width 0.2pt}
	\hrule height 0.2pt}}}$
               \ \ \ \ \ \ }
         \bigskip}
\def\real{{\bf R}}

\def\natural{{\bf N}}

\def\integer{{\bf Z}}

\def\PROOF{\medskip\noindent{\bf Proof.\ }}
\def\REMARK{\medskip\noindent{\bf Remark.\ }}
\def\LIKEREMARK#1{\medskip\noindent{\bf #1.\ }}
%%%%%%%%%%%%%%%%%%%%%%%%%%%%%%%%%%%%%%%%%%%%%%%%
% choice of default layout %%%%%%%%%%%%%%%%%%%%%
\tenpoint
%%%%%%%%%%%%%%%%%%%%%%%%%%%%%%%%%%%%%%%%%%%%%%%%%%%
%%%% paragraphs ...                        %%%%%%%%
%%%%%%%%%%%%%%%%%%%%%%%%%%%%%%%%%%%%%%%%%%%%%%%%%%%
\normalbaselineskip=5.25mm
\baselineskip=5.25mm
\parskip=10pt
\beforesectionskipamount=24pt plus8pt minus8pt
\sectionskipamount=3pt plus1pt minus1pt
%\beforesectionskipamount=42pt plus5pt minus2pt
%\sectionskipamount=1truecm
%\overfullrule=0pt
%\hfuzz=2pt
\def\em{\it}
%%%%%EOF
%%%%%%%%%%%%%%%%Layout
\normalbaselineskip=12pt
\baselineskip=12pt
\parskip=0pt
\parindent=22.222pt
%\beforesectionskipamount=42pt plus5pt minus2pt
\beforesectionskipamount=24pt plus0pt minus6pt
\sectionskipamount=7pt plus3pt minus0pt
\overfullrule=0pt
\hfuzz=2pt
\nopagenumbers
\headline={\ifnum\pageno>1 {\hss\tenrm-\ \folio\ -\hss} \else {\hfill}\fi}
\font\titlefont=ptmb at 14 pt

\font\toplinefont=cmcsc10
\font\pagenumberfont=ptmb at 10pt
%%%%%%%%%%%%%%%%%%%%%%%%%%%%%%
\newdimen\itemindent\itemindent=1.5em

\def\textindent#1{\indent\llap{#1\enspace}\ignorespaces}
\def\item{\par\noindent
\hangindent\itemindent\hangafter=1\relax
\setitemmark}
\def\setitemindent#1{\setbox0=\hbox{\ignorespaces#1\unskip\enspace}%
\itemindent=\wd0\relax
\message{|\string\setitemindent: Mark width modified to hold
         |`\string#1' plus an \string\enspace\space gap. }%
}
\def\setitemmark#1{\checkitemmark{#1}%
\hbox to\itemindent{\hss#1\enspace}\ignorespaces}
\def\checkitemmark#1{\setbox0=\hbox{\enspace#1}%
\ifdim\wd0>\itemindent
   \message{|\string\item: Your mark `\string#1' is too wide. }%
\fi}
\setitemindent{3.)}
\def\SECTION#1\par{\vskip0pt plus.2\vsize\penalty-75
    \vskip0pt plus -.2\vsize
    \global\advance\SECTIONcount by 1
    \beforesectionskip\noindent
{\sectionsize\sectiontype \actualnumber.\ #1}
    \EQNcount=1
    \CLAIMcount=1
    \SUBSECTIONcount=1
    \nobreak\sectionskip\noindent}
%% defining the symbolic value
\def\ifff(#1,#2,#3){\ifundefined{#1#2}\expandafter\xdef\csname #1#2\endcsname{#3}\write16{defining #1#2}\fi}
\expandafter\xdef\csname
sintro\endcsname{1}
\expandafter\xdef\csname
egl\endcsname{(1.1)}
\expandafter\xdef\csname
epotential\endcsname{(1.2)}
\expandafter\xdef\csname
eevolOp\endcsname{(1.3)}
\expandafter\xdef\csname
figimages/classical.ps\endcsname{Fig.\ 1}
\expandafter\xdef\csname
figimages/period.ps\endcsname{Fig.\ 2}
\expandafter\xdef\csname
cpersol\endcsname{Proposition\ 1.1}
\expandafter\xdef\csname
figimages/btot.eps\endcsname{Fig.\ 3}
\expandafter\xdef\csname
figimages/ud.ps\endcsname{Fig.\ 4}
\expandafter\xdef\csname
sdilute\endcsname{2}
\expandafter\xdef\csname
ssetup\endcsname{2.1}
\expandafter\xdef\csname
cdefprob\endcsname{Definition\ 2.1}
\expandafter\xdef\csname
ekink\endcsname{(2.1)}
\expandafter\xdef\csname
elineq\endcsname{(2.2)}
\expandafter\xdef\csname
elinop\endcsname{(2.3)}
\expandafter\xdef\csname
slinearstate\endcsname{2.2}
\expandafter\xdef\csname
cL2-measure\endcsname{Definition\ 2.2}
\expandafter\xdef\csname
canderson\endcsname{Theorem\ 2.3}
\expandafter\xdef\csname
edefineNE\endcsname{(2.4)}
\expandafter\xdef\csname
cquasi-finite\endcsname{Corollary\ 2.4}
\expandafter\xdef\csname
etauvector\endcsname{(2.5)}
\expandafter\xdef\csname
ctau-number\endcsname{Lemma\ 2.5}
\expandafter\xdef\csname
cprojector\endcsname{Proposition\ 2.6}
\expandafter\xdef\csname
epositivity1\endcsname{(2.6)}
\expandafter\xdef\csname
epositivity2\endcsname{(2.7)}
\expandafter\xdef\csname
cpositivity\endcsname{Corollary\ 2.7}
\expandafter\xdef\csname
eznorm\endcsname{(2.8)}
\expandafter\xdef\csname
snonlin\endcsname{2.3}
\expandafter\xdef\csname
etubeq\endcsname{(2.9)}
\expandafter\xdef\csname
einfty-bound\endcsname{(2.10)}
\expandafter\xdef\csname
ctube\endcsname{Proposition\ 2.8}
\expandafter\xdef\csname
econe-ineq\endcsname{(2.11)}
\expandafter\xdef\csname
ccone\endcsname{Proposition\ 2.9}
\expandafter\xdef\csname
econtraction\endcsname{(2.12)}
\expandafter\xdef\csname
einvarsets\endcsname{(2.13)}
\expandafter\xdef\csname
sspeed\endcsname{2.4}
\expandafter\xdef\csname
emotion1\endcsname{(2.14)}
\expandafter\xdef\csname
ess2\endcsname{(2.15)}
\expandafter\xdef\csname
emotion2\endcsname{(2.16)}
\expandafter\xdef\csname
espeed\endcsname{(2.17)}
\expandafter\xdef\csname
cspeed\endcsname{Theorem\ 2.10}
\expandafter\xdef\csname
scollapse\endcsname{3}
\expandafter\xdef\csname
figimages/universal.ps\endcsname{Fig.\ 5}
\expandafter\xdef\csname
cuniversal\endcsname{Theorem\ 3.1}
\expandafter\xdef\csname
scoarse\endcsname{4}
\expandafter\xdef\csname
eDerrida\endcsname{(4.1)}
\expandafter\xdef\csname
eboundary\endcsname{(4.2)}
\expandafter\xdef\csname
ccollapse\endcsname{Definition\ 4.1}
\expandafter\xdef\csname
ccoarsening\endcsname{Theorem\ 4.2}
\expandafter\xdef\csname
cpositive.measure\endcsname{Proposition\ 4.3}
\expandafter\xdef\csname
cfullequation\endcsname{Corollary\ 4.4}
\expandafter\xdef\csname
smisc\endcsname{5}
\expandafter\xdef\csname
ccomparison\endcsname{Lemma\ 5.1}
\expandafter\xdef\csname
ctau-estimates\endcsname{Lemma\ 5.2}
\expandafter\xdef\csname
cg1-g2bound\endcsname{Corollary\ 5.3}
\expandafter\xdef\csname
css\endcsname{Lemma\ 5.4}
\expandafter\xdef\csname
slinearprop\endcsname{6}
\expandafter\xdef\csname
figimages/potential.ps\endcsname{Fig.\ 6}
\expandafter\xdef\csname
ePB1\endcsname{(6.1)}
\expandafter\xdef\csname
cPbound\endcsname{Lemma\ 6.1}
\expandafter\xdef\csname
emu1\endcsname{(6.2)}
\expandafter\xdef\csname
sgeometry\endcsname{7}
\expandafter\xdef\csname
edecompose1\endcsname{(7.1)}
\expandafter\xdef\csname
edecompose2\endcsname{(7.2)}
\expandafter\xdef\csname
csbound\endcsname{Lemma\ 7.1}
\expandafter\xdef\csname
cppbound\endcsname{Lemma\ 7.2}
\expandafter\xdef\csname
ebound-dt-w\endcsname{(7.3)}
\expandafter\xdef\csname
ebound-dt-g\endcsname{(7.4)}
\expandafter\xdef\csname
emotion3\endcsname{(7.5)}
\expandafter\xdef\csname
suniversal-proof\endcsname{8}
\expandafter\xdef\csname
ctwo-kinks\endcsname{Lemma\ 8.1}
\expandafter\xdef\csname
efric\endcsname{(8.1)}
\expandafter\xdef\csname
eslope\endcsname{(8.2)}
\expandafter\xdef\csname
figimages/newfront2.ps\endcsname{Fig.\ 7}
\expandafter\xdef\csname
figimages/newfront3.ps\endcsname{Fig.\ 8}
\expandafter\xdef\csname
eDuhamel\endcsname{(8.3)}
\expandafter\xdef\csname
speriodic.proof\endcsname{9}
\expandafter\xdef\csname
eclassequ\endcsname{(9.1)}
\expandafter\xdef\csname
eamplitude-integral\endcsname{(9.2)}
\expandafter\xdef\csname
figimages/elliptic.ps\endcsname{Fig.\ 9}
\expandafter\xdef\csname
eexponent-amplitude\endcsname{(9.3)}
\def\NEWDEF #1,#2,#3 {\ifff({#1},{#2},{#3})}
%%%%%%%%%%%%%%%%%%%%%%%%%%%%%%%%%%%%%%%%%%%%%%%%
\let\truett=\tt
\fontdimen3\tentt=2pt\fontdimen4\tentt=2pt
\def\tt{\hskip -5truecm\truett
\#\#\#\#\#\#\#\#\#\#\#\#\#\#  }
%\DRAFT
%%%%%%%%%%%%%%%%%%%%%%%%%%%%%%%%%%%%%%%%%%%%%%%%%%%%%%
\newcount\footcount \footcount=0
\def\footn#1{\global\advance\footcount by 1
\footnote{${}^{\number\footcount}$}{#1}}
%%%%%%%%%%%%%%%%%%%%%%%%%%%%%%%%%%%%%%%%%%%%%%%%%%%%%%
\def\FF{{\script F}}

\def\II{{\script I}}
\def\ZZ{{\script Z}}

\def\QQ{{\script Q}}
\def\SS{{\script S}}
\def\EE{{\script E}} 
\def\3HALF{\textstyle{3\over2}}
\def\DZ{{\rm D}_Z}
\def\du{\DZ  u_Z}
\def\duj(#1){\partial_{z_{#1}}u_Z}
\def\tubo{{\cal T}_{\Gamma_0,\sigma}}
\def\tub{{\cal T}_{\Gamma,\sigma}}

\let\kappa=\varkappa
\let\epsilon=\varepsilon
\let\phi =\varphi 
\let\theta\vartheta
\def\GL{\Gamma_{\rm L}}
\def\GC{\Gamma_{\rm C}}
\def\GR{\Gamma_{\rm R}}
\def\card{{\rm card}}
\def\min{{\rm min}}
\def\max{{\rm max}}
\def\ltwo{L^2(\real,d\mu)}
%**end of header
\headline={\ifnum\pageno>1 {\toplinefont Coarsening by Ginzburg-Landau}
\hfill{\pagenumberfont\folio}\fi}
{\titlefont{\centerline{Coarsening by Ginzburg-Landau Dynamics }}}
\vskip 0.5truecm
{\it{\centerline{J.-P. Eckmann${}^{1,2}$ and J. Rougemont${}^1$}}
\vskip 0.3truecm
{\eightpoint
\centerline{${}^1$D\'ept.~de Physique Th\'eorique, Universit\'e de Gen\`eve,
CH-1211 Gen\`eve 4, Switzerland}
\centerline{${}^2$Section de Math\'ematiques, Universit\'e de Gen\`eve,
CH-1211 Gen\`eve 4, Switzerland}
}}
%\vskip 0.5truecm
\vfill
{\eightpoint\narrower\baselineskip 11pt
\LIKEREMARK{Abstract}We study slowly moving
solutions of the real Ginzburg-Landau equation on the line, by a
method due to J. Carr and R.L. Pego. These are functions taking
alternately positive or negative values on large intervals. A consequence
of our approach is that we can propose a rigorous
derivation of a stochastic model of coarsening by successive
elimination of the smallest interval, which was described in earlier
work by A.J. Bray, B. Derrida and C. Godr\`eche. }
%\vfill
\eject
\tenrm

%%%%%%%%%%%%%%%%%%%%%%%%%%%%%%%%%%%
\SECT(intro)Introduction
%%%%%%%%%%%%%%%%%%%%%%%%%%%%%%%%%%%

In a series of papers ([CP1,CP2]), Carr and Pego studied the evolution
of multi-kink
initial data of the 
real Ginzburg-Landau equation:
$$\partial_t v\,=\,\partial_x^2 v+v-v^3\,=\,\partial_x^2 v + V'(v)~,
\EQ(gl) $$
with $v(x,t)\,:\,\real\times\real^+\rightarrow\real.$ 
These data are for most $x$ very close to the stationary values $v=\pm
1$
with transitions from $\pm1 $ to $\mp 1$ at certain points. We call
these points `kinks'.
Since an isolated kink moves to the stable stationary solution $\tanh
(x-x_0)$
for some $x_0\in \real$, one expects the dynamics of multi-kink data
to be slow. Carr and Pego showed that the dynamics of the position of
the kinks can be approximated by a simple potential model, with an
exponentially decaying attraction between the kinks. They attract
each other weakly, and eventually two opposite kinks collide and `annihilate'.
Carr and Pego proved their results for 
the equation on an interval subject to Neumann boundary conditions.

In this paper, we extend their method to prove similar results for the 
equation on the infinite line.
First, we show that if the kinks are initially widely separated, then
their annihilation goes through a {\em universal} sequence of shapes.
One can then ask (and answer) some
questions about the evolution of initial data with an infinity of
kinks.
Controlling this situation leads to a rigorous derivation of the
stochastic model of coarsening studied by Bray {\it et al.}\ in
[BDG,BD]. There are several interesting technical points in this
derivation, for example the problem to show that there can be no
`conspiracy' between kinks which are far apart to change the basic
potential evolution between neighboring kinks. Our analysis goes some
way towards formulating a fully probabilistic set of `reasonable'
initial conditions. However, we fall short of finding a set of
positive measure on the full line of such conditions, but we at least
can show that there are local conditions which guarantee the control
of the evolution for all times.

Our results hold for equations which are somewhat more
general than the Ginzburg-Landau equation, which derives from a
potential $V$:
If we introduce the notation
$$ V(z)\,=\,{z^2\over 2}-{z^4\over 4}~, 
\EQ(potential)
$$
then the r.h.s.~of Eq.\equ(gl) is
$$\LL (v)\,=\,\partial_x^2 v+V'(v)~. 
\EQ(evolOp) 
$$

We can extend our results to all potentials $V\in{\cal C}^3$ which satisfy:
$$
\displaylines{
V'(\pm 1)\,=\,V'(0)\,=\,0~,\cr
V(+x)\,=\,V(-x)~,\cr
V''(\pm 1)\,<0~,\quad V''(0)\,\ge\,1~,\cr
V'(x)\,\ne\, 0~{\rm for}~x\not\in\{\pm 1,\,0\}~.
}
$$
\REMARK The condition $V(x)=V(-x)$ can be replaced by the simpler
$V(1)=V(-1)$, but then the notation gets 
somewhat more involved.
We have fixed the scale by imposing $V'(\pm1)=0$. Furthermore,
to simplify the choice of cutoff functions, we have required
$V''(0)\ge1$, and this could be generalized easily to $V''(0)>0$.

We start by listing all the bounded stationary---{\it i.e.,} 
time-independent---solutions of \equ(gl). They
can be interpreted as trajectories of a free point
particle moving in the potential $V$ without friction, with $x$ being the
`time' variable (see \fig(images/classical.ps)). Note that this is an
integrable Hamiltonian system. The stationary solutions are:
%\figurewithtexplus images/classical.ps
%images/classical.tex 5.5 6.0 0.8 The mechanical interpretation of the
%equation $\LL(v)=0$.\cr  
\item{---}Three constant solutions,
$$u_\pm(x)\,=\,\pm 1~,\;\; u_0(x)=0~.$$
\item{---}Two heteroclinic solutions, which we will denote by $\psi(x)$ and
$\psi(-x)$. We fix the notation by imposing $x\psi(x)\le 0$. For $V$ given by
Eq.\equ(potential), such a solution is:
$$
\psi(x)\,=\,-\tanh\Bigl({x\over \sqrt{2}}\Bigr)~.
$$
\item{---}Periodic solutions, which are described in the following
proposition (see \fig(images/period.ps)).
%\figurewithtexplus images/period.ps images/period.tex 8.5 10.0 -1.0
%A periodic solution $\phi_P(x)$.\cr

\CLAIM Proposition(persol) For every $V$ as described above there is a
$P_0>0$ such that
for every $P\in (P_0,\infty)$, there exists a periodic
stationary solution $\phi_P(x)$ of \equ(gl) with period $2P$ and
amplitude $A$, and $\phi _P(x)$ has exactly one maximum and one
minimum per period.
Furthermore, $\phi_P\in{\cal C}^\infty $ and there is a real analytic
bijection between the 
amplitude $A\in(0,1)$ and $P=P(A)$. 
We shall assume $\phi_P(0)=\phi_P(P)=0$, $\phi_P(x)<0$
for $x\in(0,P)$.

\REMARK In fact, $P_0=V''(0)\pi$. The result stated in this
proposition is certainly not new, but in order to stay self-contained, we
give a proof in \sec(periodic.proof).

%\figurewithtexplus images/btot.eps images/btot.tex 12.0 14.0 0.5 Numerical
%simulations. Top left: time 1 to 2.7, top right: time 25 to 29.2,
%middle left: time 29.2 to 31.3, middle right: time 3668 to 3675,
%bottom left: time 5043 to 5050 and bottom right: time 5266 to 5269.\cr

\REMARK The above list is exhaustive. This can be seen by examining
\fig(images/classical.ps). The initial position $-A$ of the particle
must satisfy $A\in[0,1]$ to keep the orbit bounded. But for any such
initial condition, we have exhibited a solution of the equation
$\LL(v)=0$. Thus, there are no other bounded stationary solutions of the
equation \equ(gl).

These periodic solutions will play an important role in the
sequel.
Our aim is to show the existence of
`metastable' states, {\it i.e.}, states that are unstable, but which
creep for a very long time (see the numerical simulations
in \fig(images/btot.eps)). It is common knowledge that the solutions
$u_\pm$ are
stable, that $\psi$ is stable up to an eigenvalue 0 (corresponding to
translations) and that all the other stationary solutions are unstable. 
We want to
study the evolution of initial conditions $v(t=0)$ which are like
`crenelations': 
We define the set $Z$ of zeros of $v(t=0)$ as
$$
Z\,=\,\{ z_j\in\real, j\in\integer, z_j<z_{j+1} {\rm ~and~}
v(x=z_j,t=0)=0 {\rm ~for~all~} j \}~.
$$
We assume $v(x,t=0)$ is positive for $z_{2j}<x<z_{2j+1}$ and negative
for
$z_{2j-1}<x<z_{2j}$.
We also introduce the interval lengths $\ell_j$ defined by:
$$
\ell_i\,=\,z_i-z_{i-1}~.
$$

%These definitions generalize easily to the cases when there are no
%zeros for very negative or very positive $x\in\real$, in which case
%the index set $i\in\integer$ has to be modified.

\LIKEREMARK{Definition}A function which is of the
form above will be called {\em admissible}, and will be denoted $u_Z$.

If we look only at the zeros of the solution of Eq.\equ(gl), then we
have a reduced system of equations, for the {\em positions} of the
zeros. Thus $Z$ becomes a function of time. One of the difficulties
in the infinite domain is to show that there are `interesting'
admissible initial data {\em which remain admissible for all times}
when evolved with the Eq.\equ(gl).

The evolution of these initial data will look as follows.
First,
the positive (negative) part of $u$ approaches rapidly $+1$ ($-1$)
and domain walls form in between, which (locally) look like
$\pm\tanh(x/\sqrt{2})$ (generally, the heteroclinic
solutions). Intuitively, $\pm 1$
are stable fixed points, but the domain walls will move. Since there is no
reason for $+1$ to be preferred to $-1$ or vice-versa, the speed of
the motion of a domain wall will, to first approximation, only depend
on the sizes $\ell_i,\ell_{i+1}$ of the two domains adjacent to this wall.
Carr and Pego showed, in the finite domain,
that the speed of motion of the $i^{\rm th}$ kink is roughly
$e^{-\ell_{i+1}}-e^{-\ell_i}$. 

We follow the method of Carr and Pego to prove similar results in the
infinite domain: to a prescribed set $Z$, we
associate a function $u_Z^{(0)}$ which has 
$Z$ as the set of its zeros. In each interval $(z_i,z_{i+1})$, we set
$u_Z^{(0)}(x)$ equal to a translate of $\phi_P$ with  $P=z_{i+1}-z_i$, 
so that $u_Z^{(0)}$ is a 
continuous function, alternatively positive and negative between 
successive zeros. Then
we slightly deform this non-differentiable function near each zero to get a
smooth function $u_Z$. (The idea of gluing near the zeros
instead of gluing in the middle of the intervals was already present
in Carr and Pego and is very fruitful.) 
This function $u_Z$ is, by construction,
equal to a stationary solution of \equ(gl), except near the set $Z$.
The next
step is the study of the stability of these `almost stationary'
functions. We show that the unstable directions are approximately
tangent to $\MM=\{u_Z\,:\,Z $ in some restricted set $ \Omega_\Gamma\}$ and 
the spectrum of the linearized operator corresponding to these
unstable modes is contained in a ball of radius $\sup_ie^{-\ell_i}$. 
%\figurewithtexplus images/ud.ps images/ud.tex 8.7 10.0 -1.2 The function
%$u_Z$. The zeros are labeled $z_i$. Bold lines indicate
%the regions of gluing.\cr

In the next section, we analyze the behavior of initial conditions close
to $u_Z$, when all kinks are far apart, in particular, we show the
existence of an invariant
neighborhood of $\MM$. In \sec(speed), we provide an explicit formula
for the speed of the kinks. In \sec(collapse), we discuss the
annihilation of a pair of neighboring kinks.
This analysis yields
a version of the Bray-Derrida-Godr\`eche dynamics of intervals which 
continuously eliminates the smallest interval, replacing it by the
union of its
neighbors ([BDG]). In \sec(coarse), we construct a set $\CC$ of
initial conditions which never come to rest,
{\it i.e.,} that `coarsen' forever in the sense introduced in [BDG].
Most of the proofs are given in \sec(misc) to \sec(periodic.proof). 

\LIKEREMARK{Acknowledgments}We are grateful to P. Collet for suggesting
this problem to us. We also thank C.-A. Pillet,
P. Wittwer, and L. Rey-Bellet for useful comments. This work was
supported by the Fonds National Suisse.

%%%%%%%%%%%%%%%%%%%%%%%%%%%%%%%%%%%%%%%%%%%%%%%%%%%%%%%%%%%
\SECT(dilute)Dynamics in the dilute state

In this section, we give the basic definitions used throughout the
paper and analyze the behavior of a well-separated collection of kinks.

\SUBSECT(setup)Definitions

Let $Z=\{z_j\}_{j\in\integer}\in\real^\integer$ be a sequence of
positive real numbers. Let 
$$\eqalign{
\ell_j\,&=\,z_j-z_{j-1}~,\cr
|Z|\,&=\,\inf_{j\in\integer}\ell_j~,\cr
c_j\,&=\,\HALF(z_j+z_{j-1})~.
}
$$
Let $\Gamma>P_0$ and suppose that $|Z|>\Gamma$. In particular, this
means that $z_{j+1}>z_j$ for all $j\in\integer$. Let  $\Omega _\Gamma$
denote the set of such $Z$:
$$\Omega_\Gamma=\left\{Z=\{\dots,z_{-1},z_0,z_1,\dots\}\in\real^\integer:\,
z_{j+1}-z_j>\Gamma,\,j\in\integer\right\}~.$$

We equip $\Omega_\Gamma$ with a probability measure $P$:

\CLAIM Definition(defprob)
Let $\{\ell_j\}_{j\in\integer}$ be i.i.d.\ random variables with a
probability density $\rho_\Gamma(x),\,x\in\real^+,$ satisfying
$\rho_\Gamma(x)>0$ for $x>\Gamma$ and $\rho_\Gamma(x)=0$ for $x\le\Gamma$.
The probability measure $P$ on $\Omega_\Gamma$ is then induced by 
choosing, for $Z\in\Omega_\Gamma$, 
$$
z_0\,=\,0~,\quad
z_j-z_{j-1}\,=\,\ell_j~.
$$

For $Z\in\Omega_\Gamma$, we construct the function $u_Z(x)$ as
described in the introduction: Let 
$$
u^{(j)}(x)\,=\,\cases{
\bigl(1-\Delta(x-z_j)\bigr
)\phi_{\ell_j}(x-z_{j-1})+\Delta(x-z_j)\phi_{\ell_{j+1}}
(x-z_{j+1})~,&\cr
\hfill{\rm if~}
j {\rm ~is~even}~,&\cr
{\bigl(1-\Delta(x-z_{j})\bigr
)\phi_{\ell_{j}}(x-z_{j})+\Delta(x-z_{j})\phi_{\ell_{j+1}}
(x-z_{j})}~,&\cr
\hfill{\rm if~}
j {\rm ~is~odd}~,&\cr
}
$$ 
where  $\Delta(x)$ is a ${\cal C}^\infty$ monotone cutoff
function satisfying:
$$\Delta(x)\,=\,\cases{0 &if $x\le -1 $~,\cr
		1 &if $x\ge 1$~.\cr}$$ 
Then, $u_Z(x)$ is given by the formula
$$
u_Z(x)=\sum_{j=-\infty}^{\infty} u^{(j)}(x)  {\bf 1}_j(x)~,\EQ(kink)
$$
where ${\bf 1}_j$ is the indicator function of the interval
$I_j\equiv[c_{j},c_{j+1}]$. Note that $u_Z\in{\cal C}^\infty$.

Expanding \equ(evolOp) around $u_Z$ by setting $v=w+u_Z$ gives:
$$
\LL(w+u_Z)\,=\,\LL(u_Z)-L_Z w+ w^2 r(w,u_Z)~, 
\EQ(lineq) 
$$
where the linear operator reads:
$$
L_Z w\,=\,-\partial_x^2w-V''\bigl(u_Z\bigr)w~,
\EQ(linop) 
$$
and the non-linear remainder is given by:
$$
r(f,g)\,=\,\int_0^1 ds\,(1-s)\,V'''(s f +g)~.
$$

%%%%%%%%%%%%%%%%%%%%%%%%%%%%%%%%%%%%%%%%%%%%%%
\SUBSECT(linearstate)Properties of the linear operator
%%%%%%%%%%%%%%%%%%%%%%%%%%%%%%%%%%%%%%%%%%%%%%

In this section, we present some properties of the linear operator
$L_Z$ defined in Eq.\equ(linop) acting on $\ltwo$ where
$\mu$ is the measure defined in the following:

\CLAIM Definition(L2-measure)
Let $\Lambda$ be a compact interval in $\real$ and $\epsilon >0$. Let
$\mu$ be an absolutely
continuous measure on $\real$ satisfying 
$$
\eqalign{
\mu(\Lambda)\,&=\,1-\epsilon~,\cr
\mu(\real\backslash\Lambda)\,&=\,\epsilon~,\cr
\exists C>0:\,\left |{d\mu(x)\over dx}-C\right |\,&\le\, \epsilon ~,
\quad {\rm for~}x\in\Lambda~. 
}
$$
The corresponding $\ltwo$-norm is denoted $\|\cdot\|_\Lambda$ and the scalar
product $\langle\cdot,\cdot\rangle_\Lambda$. 

We first describe the spectrum of $L_Z$:

\CLAIM Theorem(anderson)
There exist constants $c_1<\infty $, $M>0$, and a set 
$\Omega^*\subset\Omega_{|Z|}$ such that for sufficiently large $|Z|$,
\item{---}$P(\Omega^*)=1$,
\item{---}For all $Z\in\Omega^*$,
the $\ltwo$-spectrum of $L_Z\equiv-\partial_x^2 -
V''(u_Z)$ in $(-\alpha ,\alpha)$, with $\alpha =\OO(e^{-c_1|Z|})$,
is pure point with exponentially decaying eigenfunctions $e_j$,
$j\in\natural$. The remainder of the spectrum is contained in
$[M,\infty )$.

The proof of this result essentially follows the lines of [FSW] or
[S], and is sketched in \sec(linearprop). There is a corollary to this
theorem:

\CLAIM Corollary(quasi-finite)
Let $\epsilon >0$ and $\Lambda\subset\real$ be a compact
interval. Then there exists an $N_\epsilon <\infty $ such that for all
$w\in\ltwo\cap L^\infty $, 
$$\sum_{j>N_\epsilon}|\langle e_j(x),w\rangle_\Lambda|\,\le\,\epsilon ~.
\EQ(defineNE)
$$

\PROOF The l.h.s.\ of  Eq.\equ(defineNE) is the projection of $w$ onto
a space of functions which have exponentially small tails in $\Lambda
$. Furthermore, $w\in L^\infty $, hence $\|{\bf 1}_{\real\backslash\Lambda}
w\|_\Lambda\le\epsilon $, where ${\bf 1}_{\real\backslash\Lambda}$ is
the indicator function of the complement of $\Lambda $. 
\QED 

We now define vectors in $\ltwo$ which `generate' the translation of the
$j^{\rm th}$ kink:
$$\tau_{z_j}\,=\,(-1)^j\Theta_j(x)\partial_xu_Z(x)~,\EQ(tauvector)$$
where $\Theta_j$ 
is a (smooth) characteristic function of the interval
$I_j$: $\Theta_j\in{\cal C}^\infty$,
$$\Theta_j(x)=\cases{0~,&if
 dist$(x,I_j)>1$~,\cr
1~,&if
$x\in I_j$~,\cr
}
$$
in such a way that all its derivatives are uniformly bounded in $j$.

\CLAIM Lemma(tau-number)
For $|Z|$ sufficiently large and for $\epsilon >0$, there is a
$D_\tau$, $0<D_\tau<\infty $, such that for all $k>D_\tau$, one has
$$
\|\tau_{z_k}\|_\Lambda\,\le\,\epsilon ~.
$$ 

\PROOF The claim follows immediately from \clm(L2-measure) and from
the fact that $\tau_{z_j}$ has compact support and is uniformly bounded.
\QED

We denote by $P_{N_\epsilon}:\ltwo\to\bigoplus_{j\le N_\epsilon}\HH_{\lambda_j}$ the
spectral projector associated with the eigenvalues
$\lambda_1,\dots,\lambda_{N_\epsilon}$ of $L_Z$ (and  $\HH_{\lambda_j}\subset
\ltwo$ the corresponding spectral
subspaces). Then, if $w\in\ltwo$ satisfies $\langle
w,\,\tau_{z_j}\rangle_\Lambda=0$ for all $|j|\le D_\tau$, then its projection
onto $\bigoplus_{j\le N_\epsilon}\HH_{\lambda_j}$ is small:

\CLAIM Proposition(projector)
Let $w\in\ltwo$. There
exist constants $c_1>0$, $c_2>0$, and $D_\tau>0$ such that for
sufficiently large $|Z|$, if $w$
satisfies $\langle w,\,\tau_{z_j}\rangle_\Lambda=0$ for all $|j|\le D_\tau$,
then $\|P_{N_\epsilon}w\|_\Lambda\le c_2e^{-c_1|Z|}\|w\|_\Lambda$.

The proof of this statement can be found in \sec(linearprop).

This proposition basically says that a function $w$ which is
orthogonal to vectors $\tau_{z_j}$ located in $\Lambda$ is also
almost orthogonal to relevant unstable modes. We can infer (see
\sec(linearprop))
the following corollary: 

\CLAIM Corollary(positivity)
Let $w\in\ltwo$. Then, for sufficiently large $|Z|$, there
exist constants $M_1>0$, $M_2>0$, and $M_3>0$ such that if $w$
satisfies $\langle w,\,\tau_{z_j}\rangle_\Lambda=0$ for $|j|\le D_\tau$,
then one has:
$$
\|L_Zw\|_\Lambda^2\,\ge\,M_2\,\langle w,\,L_Zw\rangle_\Lambda
\,\ge\,M_1M_2\,\|w\|_\Lambda^2~,\EQ(positivity1)
$$
and denoting by $\chi_\Lambda$ a smooth characteristic function of the
interval $\Lambda$ ({\it i.e.,} $\chi_\Lambda(x)=1$ if $x\in\Lambda$,
$\chi_\Lambda$ has
compact support and $|\partial_x\chi_\Lambda(x)|<\HALF$ for all $x$), one has:
$$\|\chi_\Lambda w\|_\infty^2 \,\le\,M_3\langle
w,L_Zw\rangle_\Lambda~.\EQ(positivity2)$$

Motivated by the above statement, we can introduce the following norm
for perturbations $w\in\ltwo$
which are orthogonal to ${\rm span}\{\tau_{z_j}\}$, $j\in\integer$:
$$\|w\|_Z^2\equiv\langle w,\,L_Zw\rangle_\Lambda~,\EQ(znorm)$$
where $Z\in\Omega_\Gamma$. We will use this last corollary as follows:
In the next section, we construct a decomposition of a solution $v_t$
of Eq.\equ(gl) as $v_t=u_{Z_t}+w_t$ with $Z_t\in\Omega_\Gamma$ and
$w_t$ satisfying the hypothesis of \clm(positivity). Hence $w_t$
essentially decays with rate $M_2$ and we only have to work out
the evolution of $Z_t$.

%%%%%%%%%%%%%%%%%%%%%%%%%%%%%%%%%%%%%%%%%%%%%%
\SUBSECT(nonlin)Geometric structure  
%%%%%%%%%%%%%%%%%%%%%%%%%%%%%%%%%%%%%%%%%%%%%%

We next introduce the space of initial conditions for the dynamics
given by Eq.\equ(gl): Let 
$$
\tub\,=\,\left \{v\in L^\infty(\real) : \|v\|_\infty \le 1,\, 
\inf\limits_{Z\in\Omega_\Gamma}\|\chi_\Lambda( v-u_Z)\|_\infty <\sigma,\,
\inf\limits_{Z\in\Omega_\Gamma}\|v-u_Z\|_\Lambda<\infty\right\}~,\EQ(tubeq)
$$
with $\chi_\Lambda$ as in \clm(positivity).

\REMARK It is well known [G] that if $\|v_0\|_\infty \le 1$ then the
corresponding solution $v_t$ of Eq.\equ(gl) satisfies $\|v_t\|_\infty
\le 1$ for all times $t>0$. Thus, if $v_t\in\tub$, we have, for some
$Z\in\Omega_\Gamma$, 
$$
\|v_t-u_Z\|_\Lambda\,\le\,\|\chi_\Lambda(v_t-u_Z)\|_\infty
+\epsilon\|(1-\chi_\Lambda)(v_t-u_Z)\|_\infty
\,\le\,\sigma+2\epsilon~.\EQ(infty-bound)
$$

\LIKEREMARK{Important terminology}The `tube' $\tub$ depends on two 
parameters: $\Gamma$ and $\sigma$. They measure its `length' and its
`width'. Throughout this paper, we shall use
the condition `for sufficiently small $\tub$' to mean `for
sufficiently large $\Gamma<\infty$ and sufficiently small $\sigma>0$'.

\CLAIM Proposition(tube)
For sufficiently small $\tub$ and $v\in\tub$, there exists a
differentiable function $Z:\tub\rightarrow\Omega_\Gamma$ such that $\langle
v-u_{Z(v)},\tau_{z_j(v)}\rangle_\Lambda=0,\,$ for $|j|\le
D_\tau+1$. Furthermore, for all $\epsilon >0$, there is a
$Z^*\in\Omega^*$ such that
$\|(L_{Z(v)}-L_{Z^*})w\|_\Lambda<\epsilon\|w\|_\Lambda~.$

The proof of this proposition is an application of the Implicit
Function Theorem and is detailed in \sec(geometry).

\CLAIM Proposition(cone)
There exists a constant $B>0$ such that for sufficiently small $\tub$,
the following holds:
If $v_0\in\tub$, then as long as $v_t\in\tub$, one has:
$$(\partial_t+\HALF M_2)\Bigl(\|v-u_{Z(v)}\|_{Z(v)}^2-Bg_1^2(Z(v))\Bigr)
\,\le\,0~, \EQ(cone-ineq)
$$
where $M_2$ is like in \clm(positivity),
$g_1^2(Z)=\sum_{|j|\le
D_\tau}|\langle\LL(u_Z),\,\tau_{z_j}\rangle_\Lambda|^2$, 
and $\|\cdot\|_Z^2$ is given by Eq.\equ(znorm). Furthermore,
$g_1(Z)\to 0$ as $|Z|\to\infty $.

The proof of this result can be found in \sec(geometry).

This result can be converted into a contraction statement as follows:
By \clm(cone) and \clm(positivity), we have:
$$
\|w\|_{Z(v)}^2\,\le\,
Bg_1^2(Z(v))+\Bigl(
\|w\|_{Z(v_0)}^2-Ag_1^2(Z(v_0))\Bigr)e^{-M_2 t/2}~,\EQ(contraction)
$$ 
using Gronwall's Lemma. Choosing a number
$s$ in the set
$\{s\in\real^+:s^2+B\sup_{Z\in\Omega_\Gamma}g_1^2(Z)<M_2\sigma^2\}$
which is not empty for $|Z|$ sufficiently large, we can define the
following two sets:
$$\eqalign{\AA\,&=\,\bigl\{v=w+u_Z\in\tub\,:\,\|w\|_Z<s\bigr\}~,\cr
\ZZ\,&=\,\bigl\{v\in\AA\,:\,\|w\|_Z^2<Bg_1^2(Z)\bigr\}~.
}\EQ(invarsets)
$$
By Eq.\equ(contraction), we see that $\AA$ is exponentially attracted
towards $\ZZ$, and \clm(positivity) implies that $\AA\in\tub$. 
Denoting $v_t\equiv v(\cdot,t)$ a solution of
Eq.\equ(gl), we see that for $v_0\in\ZZ$, as long as $|Z(v_t)|>\Gamma$,
then $v_t\in\ZZ$. Hence, the only way to leave $\ZZ$ is to reach the
boundary $|Z|=\Gamma$. 

%%%%%%%%%%%%%%%%%%%%%%%%%%%%%%%%%
\SUBSECT(speed)Speed of the walls
%%%%%%%%%%%%%%%%%%%%%%%%%%%%%%%%%

We want to write equations for the time evolution of the function
$Z(t)\equiv Z(v_t)$ where $v_t$ is the solution of Eq.\equ(gl) with an
initial condition $v_0\in\ZZ$ and for $t<\sup\{t:|Z(v_t)|>\Gamma\}$.

Note first that
$$
\partial_t u_{Z(v)}\,=\,\sum_{j\in\integer}\left .\bigl
(\partial_{z_j}u_Z\bigr )\right |_{Z=Z(v)}\dot z_j~,
$$
where
$$
\dot z_j\,=\, \partial_t z_j(v(t))~.
$$
We shall use the more compact notation
$$
\partial_t u_{Z(v)}\,=\, \du \cdot \partial_t Z(v) ~.
$$
Differentiating the identities
$$
\langle v-u_{Z(v)},\,\left . (\partial_{z_j} u_Z)\right
|_{Z=Z(v)}\rangle_\Lambda\,=\,0~,\quad{\rm for~}|j|\le D_\tau+1~,
$$
with respect to $t$, we get, for $|j|\le D_\tau+1$ and $w=v-u_{Z(v)}$:
$$\eqalign{
\langle\LL (v),\,\tau_{z_j(v)}\rangle_\Lambda\,&=\,
\langle\du \cdot\partial_t Z(v) ,\,\tau_{z_j(v)}\rangle_\Lambda
-\langle v-u_{Z(v)},\,\DZ\tau_{z_j(v)}\cdot\partial_t
Z(v)\rangle_\Lambda~,\cr
\partial_t w\,&=\,\LL(v)- \du\cdot\partial_t Z(v) ~.}
\EQ(motion1) 
$$
We define the matrix
$$
\eqalign{
\tilde\SS\,&=\,\Bigl (\tilde\SS_{ij}\Bigr )_{-D_\tau-1\le i,j\le D_\tau+1}\cr
\,&=\,\Bigl
(\langle\partial_{z_j}u_{Z(v)},\,\tau_{z_i(v)}\rangle_\Lambda \,-\,\langle
v-u_{Z(v)},\,\partial_ {z_j(v)}\tau_{z_i(v)}\rangle_\Lambda
\Bigr)_{-D_\tau-1\le i,j\le D_\tau+1}~.
}
$$
We write Eq.\equ(motion1) in the following matrix notation:
$$
\tilde\SS\cdot\dot Z\,=\,\langle \LL(v),\tau_Z\rangle_\Lambda+\dot Z\delta_{|j|,D_\tau+1}\OO(\epsilon )~,
$$
where $\delta _{i,j}$ is the Kronecker delta function. We introduce
the notation 
$$
\SS\,=\,\tilde\SS-\delta_{|j|,D_\tau+1}\OO(\epsilon )~.\EQ(ss2)
$$ 
It will be proved in \sec(misc) that the matrix $\SS$ is invertible. Hence
the equations \equ(motion1) become
$$\eqalign{
\partial_tZ(v)\,&=\,\SS^{-1}\cdot\langle\LL(v),\,\tau_{Z(v)}
\rangle_\Lambda~,\cr
\partial_t w\,&=\,\LL(w+u_{Z(v)})-\du\cdot\SS^{-1}\cdot\langle\LL(v),\,
\tau_{Z(v)}\rangle_\Lambda~.}\EQ(motion2) 
$$

\CLAIM Theorem(speed)
There exist $c_1>0$ and $E>0$ such that for sufficiently small $\tub$,
$v_t\in\ZZ$ and $Z=Z(v_t)$, one has:
$$
\partial_tz_j\,=\, E\Bigl
(e^{-c_1\ell_{j+1}}-e^{-c_1\ell_j}\Bigr)+\OO\Bigl(
e^{-c_1\inf_{j\in\integer}\ell_j}\sup_{j\in\integer}\bigl 
(e^{-c_1\ell_{j+1}}-e^{-c_1\ell_j}\bigr)\Bigr
)+\OO(\epsilon )~,\quad z_j\in\Lambda~.\EQ(speed)
$$

The proof of this theorem is provided in \sec(geometry).

%%%%%%%%%%%%%%%%%%%%%%%%%%%%%%%%%%%%%%%%%%%%%%%%%%%%%%%%%%
\SECT(collapse)Collapse of a domain
%%%%%%%%%%%%%%%%%%%%%%%%%%%%%%%%%%%%%%%%%%%%%%%%%%%%%%%%%%

The discussion so far followed Carr and Pego quite closely. Now,
we are going to use the freedom of working with an infinite line
to get a more precise description of the collapsing mechanism. This is
possible because {\em any} distribution of kinks which is sufficiently
`dilute' and does not get stuck inside $\tub$ for all times, leaves
$\tub$ through the `needle hole' at
the $\Gamma$-end of the tube. This leads to an almost universal shape
of the solution in the interval $I_j$ which has length $\Gamma$, under
the hypothesis that it was sufficiently large at start (larger than
$\Gamma_0\gg\Gamma$).
This is illustrated by numerical integration in
\fig(images/universal.ps). 
Once the universal shape is reached, the kinks will collapse in a time
$T_p<\infty$, and the function will have constant sign in
the interval $I_j$.

%\figurewithtexplus images/universal.ps images/universal.tex 6.5 8.0 0.2
%Two different initial conditions leading to the same shape just before 
%the collapse.\cr

We next give a precise description of this final stage. Suppose
that $v_t\in\ZZ$ for all $t<T$ and that $Z=Z(v_T)$ satisfies
$|Z|=\Gamma$. Let
$w_t=v_t-u_{Z(v_t)}$. Then, by \clm(projector) and \clm(quasi-finite), 
$$
\|w_T\|_\Lambda\,\le\,\|P_{N_\epsilon}w_T\|_\Lambda+\|(1-P_{N_\epsilon})w_T\|_\Lambda\,\le\,
c_2\bigl (e^{-c_1\Gamma}+\epsilon +e^{-M T}\|w_0\|_\Lambda)~.
$$
By the definition of $\ZZ$ and by \clm(cone), $\|w_0\|_\Lambda\le Bg_1(Z)\le
c_2e^{-c_1\Gamma_0}$. Hence $\|w_T\|_\Lambda\le\Lambda
(\epsilon+e^{-c_1\Gamma})$ (see \clm(g1-g2bound) below).
In the following theorem, we study the behavior of $v_0=u_Z$ where $Z$
satisfies: There is a $j\in\integer$ such that $\ell_j=\Gamma$,
$\ell_{j\pm1}>\Gamma_0$. In \sec(universal-proof) we will prove the
following theorem:

\CLAIM Theorem(universal)
For sufficiently small $\tub$ and $\tubo$, with $\Gamma_0>\Gamma$, the
following holds: Let $v_0\in\tubo$.
Suppose that for some $T>0$ and some $i\in\integer$ one has
$z_{i+1}(v_T)-z_{i}(v_T)=\Gamma$, and all other
$z_{j+1}(v_T)-z_j(v_T)>\Gamma_0$. Then 
there is a finite $T_p$ such that
$\lim_{t\uparrow T_p}z_{j+1}(v_{t+T})-z_j(v_{t+T})=0$. 

\REMARK For large $\Gamma_0$ the collapsing time $T_p$ is
in fact essentially independent of $v_0$, and the local shape of the
two collapsing kinks is universal (independently of $i$).

%%%%%%%%%%%%%%%%%%%%%%%%%%%%%%%%%%%%%%%%%%%%%%%%%%%%%%%%%
\SECT(coarse)Existence of the coarsening dynamics
%%%%%%%%%%%%%%%%%%%%%%%%%%%%%%%%%%%%%%%%%%%%%%%%%%%%%%%%%%

In this section, we want to describe a probabilistic point of view on
the dynamics of the kinks. Since by the above discussion, the
Ginzburg-Landau dynamics of many-kink states is essentially specified
by the location of these kinks, we will treat a model which implements
the dynamics of the (discrete) set of interval lengths.

In the last section, we found an `effective' equation (Eq.\equ(speed)) for the
coordinates $\{z_j\}_{j\in\integer}$ of the zeros of a solution
of the Ginzburg-Landau equation (Eq.\equ(gl)). Getting rid of the constants
and neglecting higher order terms, this equation is:
$$
\dot z_j\,=\,e^{-(z_{j+1}-z_j)}-e^{-(z_j-z_{j-1})}~,
\quad {\rm for~}j\in\integer~.
$$
Passing to the variables $\ell_j=z_j-z_{j-1}$ (the
interval lengths), we obtain:
$$\dot\ell_j\,=\,e^{-\ell_{j+1}}+e^{-\ell_{j-1}}-2e^{-\ell_j}~.
$$
Introducing $\beta _j=e^{-\ell_j}$ yields:
$$\dot\beta _j\,=\,\beta _j\bigl (2\beta _j-\beta _{j+1}-\beta
_{j-1}\bigr )~.\EQ(Derrida)$$
Furthermore, we define a `boundary condition':
If there exists an index $j\in\integer$ and a time $t>0$ such that $\beta
_j(t-0)=e^{-\Gamma}$ ({\it i.e.}, $\ell_j(t-0)=\Gamma$) then 
$\beta_i(t),\,i\in\integer$ is defined by
$$\beta _i(t)\,=\,\cases{\beta _i(t-0) &if $i<j-1$~,\cr
		\beta _{j-1}(t-0)e^{-\Gamma}\beta _{j+1}(t-0) &if $i=j-1$~,\cr
		\beta _{i+2}(t-0) &if $i>j-1$~.\cr }\EQ(boundary)$$
(This corresponds to the merging of the two intervals $\ell_{j-1}$
and $\ell_{j+1}$ when $\ell_j$ vanishes.) The equations
\equ(Derrida) together with \equ(boundary) define a dynamics on the
space $\EE=[0,e^{-\Gamma}]^\integer$ which we baptize `coarsening dynamics' in
reference to the Bray-Derrida-Godr\`eche model.

\CLAIM Definition(collapse)
A collapse for $\beta(t)$ satisfying the coarsening dynamics is a time
$\tau$ such that $\beta (t)$ is discontinuous at $t=\tau $ ({\it
i.e.,} there exists an integer $j$ such that $\beta_j(\tau -0)=e^{-\Gamma}$).

We will exhibit a set $\CC$ of initial conditions in $\EE$ such that
the corresponding coarsening dynamics will collapse infinitely often. In
terms of the variables $z_j$, this set can be viewed as a subset of
$\real$. Its restriction to any
compact subset of $\real$ has positive measure with respect to the
probability measure $P$ introduced above.

\REMARK In their model, Bray, Derrida and Godr\`eche describe what
should be the asymptotic distribution of interval lengths. Our
distribution seems to favor larger intervals than what they
expect. However, there are two special features
of their model which we do not require here: they allow for only a
countable set of interval lengths ($\ell\in\natural$) and they study
configurations of finite volume (they take finitely many
intervals and then study a scale invariant limit, which is maybe
equivalent to taking the limit of infinitely many intervals). However,
qualitatively, our results are similar to theirs. It would be nice to
be able to show that there is a set of
initial configurations for Eq.\equ(gl) of positive measure, such that
the evolution does not tend to a stationary state. However, one should
keep in mind that the periodic solutions (\clm(persol)) have stable
manifolds. Although these manifolds should be of `measure zero', we
cannot explicitly construct them, and because of the way we estimate
the evolution, it is not even obvious to construct a set of initial
conditions which is guaranteed not to intersect these
manifolds. Controlling the evolution for larger and larger times makes
the measure of this set shrink to zero. One should also keep in mind that
$P$ is not a measure on the space of initial conditions for
Eq.\equ(gl) but on the space of the positions of the zeros of such an
initial condition. There are really many functions whose set of zeros is an
element of $\CC$, and the analysis of the preceding sections shows
that (almost) all such functions have the same long-time
behavior. Any initial configuration in $\ZZ$ for the dynamics of
Eq.\equ(gl) defines an element of $\CC$. The evolution in $\ZZ$ can
then be reconstructed from the corresponding evolution in $\CC$ under
Eq.\equ(Derrida), up to terms of order $e^{-\Gamma}$ and terms of order
$\epsilon $, via the function $u_Z$ given by Eq.\equ(kink). 

Let us define the set $\CC$:
$$\CC\,=\,\Bigl\{\beta\in\EE\,:\,\exists\{j_n\}_{n\in\natural}\subset\integer
{\rm ~s.t.~}\forall n\in\natural, ~\beta _{j_n}\in({\textstyle{e^{-\Gamma}\over
2n}} , {\textstyle{e^{-\Gamma}\over n}}),\;\beta _{j_n\pm
1}\in(0 , {\textstyle{e^{-\Gamma}\over 6n}})\Bigr\}~,$$ 
and state the result:

\CLAIM Theorem(coarsening)
Let $t\rightarrow\beta (t)\in\EE$ be the coarsening dynamics
associated with an
initial condition $\beta (0)\in\CC$. Then there
exists an infinite sequence of numbers $0<\tau _1<\tau _2<\dots
<\tau _n<\dots$,
such that $\tau _m\rightarrow\infty$ when $m\to\infty$, and, for all
$n\in\natural$,
$\tau_n$ is a collapse for $\beta (t)$.

\PROOF
Suppose $\beta _0(0)\in ({e^{-\Gamma}\over 2n},{e^{-\Gamma}\over n})$
and $\beta _{\pm 1}(0)<e^{-\Gamma}/(6n)$. Then, by \equ(Derrida)
$$
\eqalign{
\dot\beta _0(0)\,&\ge\,{\textstyle{2e^{-\Gamma}\over 3n}}\beta _0(0)~,\cr
\dot\beta _{\pm 1}(0)\,&\le\,\beta _{\pm 1}\bigl (2\beta _{\pm
1}-\beta _0\bigr )\,<\,0~.}
$$
In addition, we note that $\dot\beta _0(t)\le 2\beta _0(t)$.
Hence $\beta _{\pm
1}(t)<e^{-\Gamma}/(6n)$, which implies 
$$
{\textstyle{e^{-\Gamma}\over
2n}}e^{2e^{-\Gamma}t/(3n)}\,<\,\beta_0(t)\,<\,{\textstyle{e^{-\Gamma}\over
n}}e^{2t}~,
$$
for all times $t<\sup\{t:\beta_0(t)<e^{-\Gamma}\}$,
from which follows that there is a time $\tau_n$ in the interval
$(\HALF\log n,\3HALF ne^\Gamma\log(2n))$ such that
$\beta_0(\tau_n)=e^\Gamma$. Hence there exists a subsequence
$\tau_{n_j}$ satisfying the claim.
Note that the fact that collapses may occur elsewhere
in the meantime is irrelevant, since (apart from shifting the indices)
it cannot modify $\beta _0$ and it can only make $\beta _{\pm 1}$ even smaller.
\QED
Taking an interval $\Lambda\subset\real$, we define $\CC^\Lambda$ as the set
$\CC$ (viewed as a subset of $\real$ through the correspondence $\{\beta
_j\}_{j\in\integer}\leftrightarrow\{z_j\}_{j\in\integer}$)
restricted to $\Lambda$.

\CLAIM Proposition(positive.measure)
Let $\Lambda$ be a compact interval in $\real$ and $|\Lambda|$ be its
length. Then there is a $\delta=\delta(|\Lambda|)>0$ such that 
$$P\bigl (\CC^\Lambda\bigr )\,\ge\,\delta~,$$
where $P(\cdot)$ is the probability measure defined in \clm(defprob).

\PROOF
Let $\beta \in\CC$ and $\{j_n\}_{n\in\natural}$ be the indices such
that $\beta _{j_n}\in ({e^{-\Gamma}\over 2n}\,,{e^{-\Gamma}\over n})$
and $\beta _{j_n\pm 1}<{e^{-\Gamma}\over n}$. Let
$M^*=\sup\bigl\{M:\sum_{n=1}^M3\Gamma+\log 2n+2\log 6n<|\Lambda|\bigr\}$.
The interval $\Lambda$ cannot contain more than the intervals
$\ell_{j_1},\dots,\ell_{j_{M^*}}$. Hence 
$$P\bigl (\CC^\Lambda\bigr
)\,\ge\,\prod_{n=1}^{M^*}\int_{\Gamma+\log n}^{\Gamma+\log 2n}\!dx\,
\rho_\Gamma (x)\biggl
(\int_{\Gamma+\log 6n}^\infty\!dy\,\rho_\Gamma (y)\biggr )^2~.$$
By hypothesis, $\rho (x)>0,\,\forall x\in(\Gamma,\infty)$, and since 
$\Lambda$ is compact, $M^*$ is finite, hence the claim is proved.
\QED

Let us define the set $\CC^*$ which is $\CC$ written in the
variables $z_j$:
$$
\CC^*\,=\,\left \{Z\in\Omega
_\Gamma\,:\,\{e^{z_j-z_{j+1}}\}_{j\in\integer}\in\CC\right \}~.
$$
We also define $\tub^*$ by replacing $\Omega_\Gamma$ by $\CC^*$ in
the definition of $\tub$ of Eq.\equ(tubeq). The subset $\ZZ^*$ is then
defined by Eq.\equ(invarsets), replacing $\tub$ by $\tub^*$. We also
denote by $z_j(v)$ the $j^{\rm th}$ zero of the function
$v$, with $z_j(v)<z_{j+1}(v)$.

\CLAIM Theorem(fullequation)
For sufficiently small $\tub^*$, for all $v_0\in\tub^*$, if $v_t$
denotes the solution of Eq.\equ(gl) associated with the initial
condition $v_0$, there
exist a sequence of times $\{t_n\}_{n\in\natural}$ and a sequence of
indices $\{j_n\}_{n\in\natural}$ such that
$\lim_{t\to t_n}\left |z_{j_n}(v_t)-z_{j_n-1}(v_t)\right |=0$, and
$\lim_{n\to\infty }t_n=\infty $.

\PROOF Denote by $\{j_n\}_{n\in\natural}$ the indices such
that $e^{-(z_{j_n}-z_{j_n-1})}\equiv e^{-\ell_{j_n}}\in ({e^{-\Gamma}\over
2n}\,,{e^{-\Gamma}\over n})$ and
$e^{-(z_{j_n\pm1}-z_{j_n\pm1-1})}\equiv e^{-\ell_{j_n\pm
1}}<{e^{-\Gamma}\over n}$. Choose a set $\{\Lambda_j\}_{j\in\integer}$
of disjoint compact intervals of $\real$, such that for all $n$, there
exists a $k$ with $[z_{j_n-1},z_{j_n}]\subset\Lambda_k$, {\it i.e.,}
each interval of length $\ell_{j_n}$ is contained in a single interval
$\Lambda_k$. Associate with each interval $\Lambda_j$ a weight $\mu_j$
as in \clm(L2-measure). Then, by
\clm(speed), the dynamics of the zeros $z_{j_n}$ is given by
Eq.\equ(speed) and their collapse is described by \clm(universal). By
\clm(coarsening), for sufficiently small $\tub^*$, there exists a
sequence of collapsing times. This proves the assertion.
\QED

%%%%%%%%%%%%%%%%%%%%%%%%%%%%%%%%%%%%%%%%%%%%%%
\SECT(misc)Miscellaneous bounds
%%%%%%%%%%%%%%%%%%%%%%%%%%%%%%%%%%%%%%%%%%%%%

We first give estimates on the behavior of the function $u_Z$ given in
Eq.\equ(kink). In particular, we show that near the set $Z$, this
function is so close to the heteroclinic solution $\psi$, that it is
almost stationary in an $L^\infty $ sense. 

\CLAIM Lemma(comparison)
There exist positive $K $ and $c_1$, such that for sufficiently
large $|Z|$, the following holds:
\item{1)}$|\psi((-1)^{j+1}(x-z_j)) - u_Z(x)|\le K
e^{-c_1\min(\ell_j,\ell_{j+1})}$, for $|x-z_j|\le1$ and for $j\in\integer$.
\item{2)}$\|\LL(u_Z)\|_\infty \le K e^{-c_1|Z|}$,

\PROOF
We first compare $\phi_P$ with $\psi$ for fixed $P$.
Let $g(x)=\psi(x)-\phi_P(x)$, $\alpha =V(\phi_P(P/2))=V(-A(P))$,
{\it cf.~}\fig(images/period.ps), 
and suppose $x\in[-P/2,P/2]$. 
If $f$ is a stationary solution of Eq.\equ(gl) then $f''+V'(f)=0$, and
thus $f''f' + V'(f)f'=0$, {\it i.e.,} $\HALF(f')^2 +V(f)$ is constant, and
taking $x^*$ with $f'(x^*)=0$, we get
$$
\HALF(f')^2(x) + V(f(x))\,=\,V(f(x^*))~.
$$
Therefore the 
derivative
of $g$ satisfies (note that for $x\in[-P/2,P/2]$, $\phi_P(x)$ is monotone):
$$\eqalign{
	|g'(x)|\,&=\,\sqrt{2}\Bigl|\sqrt{V(\psi_{\vphantom P}(\infty
	))-V(\psi(x))}-\sqrt{\alpha-V(\phi_P(x))}\,\Bigr|\cr
        \,&=\,\sqrt{2}\Bigl|\sqrt{V(\psi_{\vphantom P}(\infty
	))-V(\psi(x))}-\sqrt{-(V(\psi(\infty
	))-\alpha)+V(\psi(\infty ))-V(\phi_P(x))}\,\Bigr|\cr
	\,&\le\,\sqrt{2}\Bigl|\sqrt{V(\psi_{\vphantom P}(\infty
	))-V(\psi(x))}-\sqrt{V(\psi(\infty
	))-V(\phi_P(x))}\Bigr|+\Bigl|\sqrt{V(\psi(\infty))-\alpha}\,\Bigr|\cr
	&\,\le\, C\bigl( |\psi(x)-\phi_P(x)|+e^{-c_1P}\bigr )
\,=\, C\bigl (|g(x)|+e^{-c_1P}\bigr )~.} $$
In the third line, we have used the 
inequality $-\sqrt{-a+b}\le\sqrt{a}-\sqrt{b}$, and in the last line, the
first term comes from the differentiability of the function $V$ while
the second term is a consequence of Eq.\equ(exponent-amplitude) below.
We apply Gronwall's lemma (and $g(0)=0$) and get
$$|g(x)|\,\le\, K_1 e^{-c_1P}~, $$
and 
$$|g'(x)|\,\le\, K_2 e^{-c_1P}~. $$
Now, recalling the definition \equ(kink) of $u_Z$, we have
$u_Z(x)=\bigl(1-\Delta(x-z_j)\bigr
)\phi_{\ell_j}(x-z_{j-1})+\Delta(x-z_j)\phi_{\ell_{j+1}}
(x-z_{j+1})$ for $|x-z_j|<\inf(\ell_j,\ell_{j+1})/2$. Hence 
$$
\min(\phi_{\ell_j}(x-z_{j-1}),\phi_{\ell_{j+1}}(x-z_{j+1}))\,\le\,
u_Z(x)\,\le\,\max(\phi_{\ell_j}(x-z_{j-1}),\phi_{\ell_{j+1}}(x-z_{j+1}))~.
$$
Consequently
$$\eqalign{&|\psi((-1)^{j+1}(x-z_j))-u_Z(x)|\cr
\,&=\,\max\Bigl(
\psi((-1)^{j+1}(x-z_j))-u_Z(x),\cr
&\,\qquad\qquad u_Z(x)-\psi((-1)^{j+1}(x-z_j)) \Bigr)\cr
\,&\le\,\max\Bigl(\psi((-1)^{j+1}(x-z_j))-\min\bigl
(\phi_{\ell_j}(x-z_{j-1}),\phi_{\ell_{j+1}}(x-z_{j+1})\bigr ),\cr
&\,\qquad\qquad\max\bigl (\phi_{\ell_j}(x-z_{j-1}),\phi_{\ell_{j+1}}
(x-z_{j+1}))-\psi((-1)^{j+1}(x-z_j)) \bigr )\Bigr)\cr
\,&\le\,\max\bigl ( K_1 e^{-c_1\ell_j}, K_1
e^{-c_1\ell_{j+1}}\bigr )~,}
$$ 
which proves claim 1).

We write, for $x\in I_j$, 
$$
\LL(u_Z)\,\equiv\,\Delta''(\phi_{\ell_j}-\phi_{\ell_{j+1}})+2\Delta'
(\phi_{\ell_j}'-\phi_{\ell_{j+1}}')-G~.
$$ 
Using the fact that $\phi_P$ is a solution of $\LL(u)=0$, we have
$$G\,=\,(1-\Delta)V'(\phi_{\ell_j})+\Delta V'(\phi_{\ell_{j+1}})-V'\bigl
((1-\Delta)\phi_{\ell_j}+\Delta\phi_{\ell_{j+1}}\bigr)~. $$
We expand $G$ near 0 and look at the coefficient of $V'''(0)/2$, which
is the first non-vanishing term:
$$\eqalign{
&(1-\Delta)\phi_{\ell_j}^2+\Delta\phi_{\ell_{j+1}}^2-\bigl((1-\Delta)\phi_{\ell_j}+\Delta\phi_{\ell_{j+1}}\bigr)^2\cr
\,&=\,(1-\Delta)\phi_{\ell_j}^2+\Delta\phi_{\ell_{j+1}}^2-(1-\Delta)^2\phi_{\ell_j}^2-\Delta^2\phi_{\ell_{j+1}}^2-2(1-\Delta)
\Delta\phi_{\ell_j}\phi_{\ell_{j+1}}\cr
\,&=\,(1-\Delta)\Delta\bigl (\phi_{\ell_j}^2+\phi_{\ell_{j+1}}^2-2\phi_{\ell_j}\phi_{\ell_{j+1}}\bigr )\cr
\,&\le\,(\phi_{\ell_{j+1}}-\phi_{\ell_j})^2~.} $$
Consequently, $|G|\le\kappa_3|\phi_{\ell_{j+1}}-\phi_{\ell_j}|^2$ and thus,  using 1),
$$
|\LL(u)|\,\le\,\kappa_1 |g(x)|+\kappa_2 |g'(x)|+\kappa_3 |g(x)|^2
\,\le\, K e^{-c_1\min(\ell_j,\ell_{j+1})}~,
$$
which completes the proof of claim 2).
\QED

We next give estimates related to the vectors $\tau_{z_j}$ introduced
in Eq.\equ(tauvector).

\CLAIM Lemma(tau-estimates)
Let $Z$ in $\Omega_\Gamma$, $c_j=\HALF(z_j+z_{j-1})$, and $\tau_{z_j}$
as defined in Eq.\equ(tauvector). Then, there exist
$ K>0$, $c_1>0$, and $c_2>0$ such that for sufficiently large $\Gamma$, one
has:
$$\eqalign{\langle\LL(u_Z),\tau_{z_j}\rangle_\Lambda\,&=\,c_2\bigl(V(u_Z(c_j))-
V(u_Z(c_{j+1})\bigr)+\OO(\epsilon )~,\quad {\rm for~}z_j\in\Lambda~,\cr
\|L_Z\tau_{z_j}\|_\Lambda\,&\le\, K
e^{-c_1\min(\ell_j,\ell_{j+1})}~,{\rm for~}|j|\le D_\tau~.
}
$$

\PROOF
We compute, using \clm(L2-measure),
$$\eqalign{
&\int_\real\!d\mu\,\LL\bigl(u_Z(x)\bigr)\tau_{z_j}(x)\,=\,C\int_{c_j+1}
^{c_{j+1}-1}\!dx\,
\bigl(\partial_x^2 u_Z+V'(u_Z)\bigr)\partial_xu_Z+\OO(\epsilon )\cr
\,=\,&
C\Bigl(\HALF(\partial_x u_Z)^2+V(u_Z)\Bigr)(c_j+1)-
C\Bigl(\HALF(\partial_x u_Z)^2+V(u_Z)\Bigr)(c_{j+1}-1)+\OO(\epsilon )\cr
\,=\,&C\bigl(V(u_Z(c_{j+1})-V(u_Z(c_j))\Bigr)+\OO(\epsilon )~.
} $$
This completes the proof of the first claim. For the second one, we
use that $\tau_{z_j}$ has compact support and is equal to a stationary
solution of Eq.\equ(gl) in the interval $I_j\cap\{x:|x-z_j|>1\}$: 
$$
\eqalign{
\|L_Z\tau_{z_j}\|_\Lambda^2\,&=\,\int_{c_j-1}^{c_{j+1}+1}\!d\mu\,\left
|\partial_x^2\tau_{z_j} +V'(u_Z)\tau_{z_j}\right |^2\cr
\,&\le\,C\int_{z_j-1}^{z_j+1}\!dx\,\left
|\partial_x^2\tau_{z_j} +V'(u_Z)\tau_{z_j}\right |^2\cr
\,&\le\,\bigl (C\sup_{|x-z_j|<1}\left|\LL(u_Z)\right |\bigr )^2~,
}
$$
and we use \clm(comparison) to conclude.
\QED

\CLAIM Corollary(g1-g2bound)
Let $g_1^2(Z)=\sum_{|j|\le
D_\tau}|\langle\LL(u_Z),\,\tau_{z_j}\rangle_\Lambda|^2$
and $g_2(Z)=\sup\bigl\{
{\|L_Z\tau\|_\Lambda\over\|\tau\|_\Lambda}:\tau\in{\rm
span}\{\tau_{z_j}\, :\,|j|\le D_\tau\}
\backslash\{0\}\bigr\}$. Then, under the hypothesis of
\clm(tau-estimates), there exists $ K_1>0$, $ K_2>0$ such
that:
$$|g_1(Z)|\,\le\, K_1e^{-c_1|Z|}~,\qquad|g_2(Z)|\,\le\, K_2e^{-c_1
|Z|}~.$$

\PROOF
The first claim follows from Eq.\equ(exponent-amplitude) and the
second from the following calculation:
Let $\tau\equiv\sum_{|j|\le D_\tau}t_j\tau_{z_j}$. Then 
$$
\eqalign{
\|L_Z\tau\|_\Lambda&\,\le\,\sum_{|j|\le D_\tau}|t_j|\|L_Z\tau_{z_j}\|_\Lambda
\,\le\,\sum_{|j|\le D_\tau}|t_j|\sup_{|j|\le D_\tau}\|L_Z\tau_{z_j}\|_\Lambda~,\cr
\|\tau\|_\Lambda&\,\ge\,\inf_{|j|\le
D_\tau}\|\tau_{z_j}\|_\Lambda\sum_{|j|\le D_\tau}|t_j|~.
} $$
We finish the proof by noting that $\tau_{z_j}(x)$ is strictly positive
in $[c_j+1,c_{j+1}-1]$ hence its norm is uniformly bounded from below
for $|j|\le D_\tau$.
Then we apply \clm(tau-estimates). 
\QED

Next we prove that certain matrices used in \sec(geometry) have a bounded
inverse.

\CLAIM Lemma(ss)
For sufficiently small
$\tub$, and all $v\in\tub$ and for $N<\infty $, the matrices 
$$\eqalign{\tilde\SS\,&=\,\Bigl (\tilde\SS_{ij}\Bigr)_{i,j=1,\dots,N}\,=\,\Bigl
(\langle\partial_{z_j}u_Z,\,\tau_{z_i}\rangle_\Lambda\,-\,
\langle v-u_Z,\,\partial_{z_j}\tau_{z_i}\rangle_\Lambda\Bigr
)_{i,j=1,\dots,N} ~,\cr
\SS_1\,&=\,\Bigl ((\SS_1)_{ij}\Bigr )_{i,j=1,\dots,N}
\,=\,\Bigl (\langle\partial_{z_j}u_Z,\,\tau_{z_i}\rangle_\Lambda\Bigr
)_{i,j=1,\dots,N} ~,}
$$ 
have uniformly bounded inverse.

\PROOF We start by the following remark: since, by our
assumption on $V$, we have 
$\pi\le P_0<\Gamma$, the tangent vectors $\tau_{z_j}$ and $\tau_{z_{j+2}}$
have disjoint support. Therefore, the matrix $\tilde\SS$
is tridiagonal and we only have to control the overlap between 
$\tau_{z_j}$ and $\tau_{z_{j\pm1}}$.
To prove that $\tilde\SS$ is invertible, we show that it
is diagonally dominant, {\it i.e.},
$$\bigl|\tilde\SS_{ii}\bigr|\,>\,\sum_{j\ne i}\bigl|\tilde\SS_{ij}\bigr|~. $$
\item{1)}The diagonal terms are
$\tilde\SS_{ii}=\langle\partial_{z_i}u_Z,\,\tau_{z_i}\rangle_\Lambda -\langle
w,\,\partial_{z_i}\tau_{z_i}\rangle_\Lambda$. The first term is
uniformly 
bounded below, by Proposition 2.3 of [CP1].
In fact, it is a consequence of 
$$
|\langle\partial_{z_i}\phi_{\ell_i},\Delta_i\partial_x\phi_{\ell_i}
\rangle_\Lambda|\,\approx\,|\langle\partial_{z_i}\psi(\cdot-z_i),\partial_x
\psi(\cdot-z_i)\rangle_\Lambda|\,=\,\|\partial_x\psi\|_\Lambda^2>K~.
$$
The second term in $\tilde\SS_{ii}$ is $\OO(\sigma)$,
thus, for sufficiently small $\tub$, the whole 
expression is bounded below.
\item{2)}We next control the off-diagonal terms
$\tilde\SS_{ij}=\langle\partial_{z_j}u_Z,\,\tau_{z_i}\rangle_\Lambda -\langle
w,\,\partial_{z_j}\tau_{z_i}\rangle_\Lambda$. The first term 
is bounded by a constant which goes to zero as $\Gamma$ goes to 
infinity, see again Proposition 2.3 of [CP1] (recall that $\tau_{z_j}$
has compact support, and the overlap between $\tau_{z_j}$ and
$\partial_{z_{j\pm1}}\phi _{\ell_j}$ is very small). The second term is
treated as before. 
\item{3)}The proof for $\SS_1$ is
a special case of 1) and 2).
\QED

\REMARK Obviously, for $\epsilon $ sufficiently small, the matrix
$\SS$ defined in Eq.\equ(ss2) is also invertible.

%%%%%%%%%%%%%%%%%%%%%%%%%%%%%%%%%%%%%%%%%%%%%%%%%%%%%
\SECT(linearprop)Proofs of the properties of the linear operator
%%%%%%%%%%%%%%%%%%%%%%%%%%%%%%%%%%%%%%%%%%%%%%%%%%%%%

The first proof we provide in this section concerns the spectrum of
$L_Z=-\partial_x^2-V''(u_Z)$. 

\LIKEREMARK{Sketch of the proof of \clm(anderson)}First of all, we
show that the operator $L_Z$ with $Z=\{-z,z\}$ (and the convention
that $\phi_\ell(x-z)=\psi(x-z)$ if $\ell=\infty $) has two eigenvalues
satisfying the bounds of \clm(anderson). The function $u_Z$ (with
$\card (Z)=2$) is
positive at $|x|\gg 1$ and negative at $x=0$.
We can view $L_Z$ as a
Schr\"odinger operator with a potential 
$U(x)=-V''(u_Z(x))$, which is a symmetric double well 
(see \fig(images/potential.ps)), $U_\min<U(x)<U_\max$.
Its spectrum is made up of isolated eigenvalues in 
$(U_\min,U_\max)$ and absolutely continuous spectrum in $[U_\max,\infty)$. 
When $|Z|\rightarrow\infty$, 
then the lowest eigenvalue is degenerate, and it is (by translation)
given as the lowest eigenvalue of
$L^*=-\partial_x^2-V''(\psi)$. In this limit,
$\psi'(x)$ is an eigenfunction of $L^*$ with eigenvalue 0 (it
corresponds to the invariance under translation). This is the ground
state, since $\psi'$ is a positive function. This double
eigenvalue splits into $\lambda_-<\lambda_+$ when $|Z|<\infty$. The
proof
uses the fact that $e_-$ ($e_+$), the corresponding
eigenfunctions, are the even (odd) extensions of the ground state of
the same operator with Neumann (Dirichlet) boundary condition at $x=0,$
and the splitting is a consequence of the Dirichlet-Neumann
bracketing. Furthermore, the splitting will be exponentially small as
$|Z|\to\infty $ (note that $|\phi_P(x)-\psi(x)|\le e^{-c_1P}$, for
$|x|<P/2$, see \clm(comparison)): one has $\lambda _+-\lambda _-\le \Lambda
e^{-c_1|Z|}$ (see [RS4], p.34, example 6).

%\figurewithtexplus images/potential.ps images/potential.tex 8.0 11.0 -0.8 The 
%potential $U(x)$ together with the function $u_Z(x)$. Note that the
%minima of $U$ coincide with the zeros of $u_Z$.\cr

By similar reasoning, for any $N<\infty $, the spectrum of $L_Z$ with
Dirichlet boundary conditions on $c_{-N}$ and $c_N$ has $2N+1$
eigenvalues satisfying the bounds of \clm(anderson). 

Next, we study the spectrum of $L_Z$ on the line, with infinitely many
kinks, which can be viewed as a Schr\"odinger operator with infinitely
many potential wells, distributed with the probability $P$.
We write
$W(x)\equiv-V''\bigl(u_Z(x)\bigr)$ for this `disordered' potential. 
We define the set of intervals 
$$\II_0=\{I_{0,j}=[z_j-\Gamma/2,z_j+\Gamma/2],\,j\in\integer\}~.$$
These are intervals of length $d_0=\Gamma$ 
centered at $c_{0,j}=z_j$, {\it i.e.,} at the bottoms of the potential
wells. Then, $S_0(Z,\,\lambda)$ is the set of the intervals
in which the potential is smaller than the energy $\lambda $ plus $\HALF$:
$$
S_0(Z,\,\lambda)\,=\,\left \{I_{0,j}\in\II_0\,:\,\forall x\in I_{0,j}\, ,
W(x)\,\le\, \lambda +\HALF\right \}~.
$$
We then inductively construct a hierarchy of sets $\II_n$ containing
intervals $I_{n,j}$ centered at $c_{n,j}$ and defined as follows: 
\item{---}If $c_{n,j+1}-c_{n,j}>3d_n$ then 
$c_{n+1,j}=c_{n,j}$ and $I_{n+1,j}$ has length $d_{n+1}=2d_n$. 
\item{---}If
$c_{n,j+1}-c_{n,j}\le 3d_n$ then $c_{n+1,j}=\HALF(c_{n,j}+c_{n,j+1})$
and $I_{n+1,j}$ has length $d_{n+1}+d_n=3d_n$. 

Hence $I_{n+1,j}$ contains
at most two intervals $I_{n,j}$ and its endpoints do not belong to an
interval $I_{m,k}$ with $m<n$. After renumbering the
intervals, we obtain a set $\II_{n+1}=\{I_{n+1,j},\,j\in\integer\}$ of
intervals with centers $c_{n+1,j}$ satisfying $c_{n+1,j+1}>c_{n+1,j}$
for all integers $j$.
We call `singular' the intervals $I_{n,j}$ satisfying:
\bigskip
{\narrower\smallskip\noindent{\bf (C1)} The operator
$L_Z(I_{n,j})$, {\it i.e.}, $L_Z$  
with Dirichlet boundary conditions on $\partial I_{n,j}$, 
has an eigenvalue $\lambda_{n,j}$ 
such that $|\lambda-\lambda_{n,j}|\le 2^{-n\beta}$,\smallskip}
\bigskip
\noindent{where $\beta$ will be fixed later. The `singular sets' are
then defined as :}
$$
S_n(Z,\,\lambda)\,=\,\left \{I_{n,j}\in\II_n\,:\,I_{n,j}~{\rm
satisfies}~{\bf (C1)}\right \}~.
$$

\CLAIM Lemma(Pbound) 
Let $I_{n,j}\in\II_n$. There is a $c_1>0$ and a $Z\in\Omega_\Gamma$
such that if 
$\Lambda^{(n)}=[\Lambda^{(n)}_1,\Lambda^{(n)}_2]\subset (-\alpha,\alpha)$,
with $\alpha=\OO(e^{-c_1|Z|})$ when $|Z|\to\infty$, we have
the following estimate:
$$ 
P\Bigl({\rm spec}(L_Z(I_{n,j}))\,\cap\,\Lambda^{(n)}\ne\emptyset\Bigr)
\,\le\, (\Lambda^{(n)}_2-\Lambda^{(n)}_1)\, 2^{2n}
\,\|\rho_\Gamma\|_\infty~.
\EQ(PB1)$$

\LIKEREMARK{Proof of the lemma}We denote by 
$$
N_L(x)\,\equiv\,\card\left \{E\le x\,:\,E
{\rm ~is~an~eigenvalue~of~}L\equiv L_Z(I_{n,j})\right \}
$$
the integrated density of states
and, if $A$ is a random variable over
the space of potentials $W$, then $\EE(A)$ is its
expectation value.
The l.h.s.\ of \equ(PB1) is bounded by
$\EE\bigl ({N_L(\Lambda^{(n)}_2)-N_L(\Lambda^{(n)}_1)}\bigr )$. Obviously
$N_L(\Lambda^{(n)}_2)\ge N_L(\Lambda^{(n)}_1)\ge 0$, and, if $N_+\equiv
N_L(\Lambda^{(n)}_2)-\EE\bigl ({N_L(\Lambda^{(n)}_1)}\bigr )$ and $N_-\equiv
N_L(\Lambda^{(n)}_1)-\EE\bigl ({N_L(\Lambda^{(n)}_1)}\bigr )$, then
$$
\eqalign{
\EE\bigl({N_L(\Lambda^{(n)}_2)-N_L(\Lambda^{(n)}_1)}\bigr
)\,&=\,\EE\bigl({N_+-N_-}\bigr )\,\le\,
\EE\bigl({N_+}\bigr )\cr
\,&=\,\EE\bigl({N_+}\bigr )-\EE\bigl({N_-}\bigr )
\,=\,\EE\bigl({N_L(\Lambda^{(n)}_2)}\bigr )-\EE\bigl(
{N_L(\Lambda^{(n)}_1)}\bigr )~.}$$
Thus the l.h.s.\ of \equ(PB1) is also bounded by
$\int_{\Lambda^{(n)}_1}^{\Lambda^{(n)}_2}\!d\lambda{d\EE\bigl({N_L}\bigr
)\over
d\lambda}$. 
Note that $I_{n,j}$ contains at most $2^n$ wells and, since $N_L(\lambda)=
N_{L-\lambda }(0)$, we can replace derivatives with respect to the second 
argument by (minus) the derivatives w.r.t.\ the first one. Note also that 
$L_Z(I_{n,j})$ depends only on the interval lengths $\ell_j$,
$j=1,\dots ,2^n$ (up to a translation of the indices), which gives:
$$
\eqalign{
\,&\,\EE\bigl({N_L(\Lambda^{(n)}_2)}\bigr
)-\EE\bigl({N_L(\Lambda^{(n)}_1)}\bigr )\cr
\,&=\,
\sum_{j=1,\dots , 2^n}\int_{\Lambda^{(n)}_1}^{\Lambda^{(n)}_2}\!d\lambda
\int_\Gamma^{\infty} \!\rho_\Gamma(\ell_j)\left (-{\partial 
N_L\over\partial\ell_j}\right)\;d\ell_j
\;\prod_{k\ne j}\rho_\Gamma(\ell_k)\;d\ell_k\cr
\,&\le\,2^n(\Lambda^{(n)}_2-\Lambda^{(n)}_1)\|\rho_\Gamma\|_\infty  
\Bigl(N_L(\ell_j=\Gamma)-N_L(\ell_j=\infty)\Bigr)\cr
\,&=\,2^{2n}(\Lambda^{(n)}_2-\Lambda^{(n)}_1)\|\rho_\Gamma\|_\infty ~,
} 
$$
where we have used that there is a number $\alpha>0$ such that
$L_Z(I_{n,j})$ has only $2^n$ eigenvalues $|\lambda _j|<\alpha$ with
$\alpha=\OO(e^{-c_1|Z|})$. 
\QED
Taking $\Lambda^{(n)}=[\lambda-2^{-n\beta },\lambda+2^{-n\beta }]$ and
$\beta>4$, the sum over all $n$ 
of the r.h.s.\ of \equ(PB1) is a finite number. We can apply 
the Borel-Cantelli Lemma and conclude that the 
probability 
$$
P\Bigl(\card\bigl\{ n\,:\,{\rm spec}(L_Z(I_{n,j}))
\cap\Lambda^{(n)}\ne\emptyset\bigr\}\,=\,\infty\Bigr)
$$ is zero, or, in
other words, that there exists almost surely a number
$N=N(Z,\lambda)<\infty$
such that all the $I_{N,j}$ violate the condition {\bf (C1)}.

The remainder of the proof is very similar to the one in [FSW], namely, 
one shows that the Green's function $G(\lambda,x,y)$ is exponentially
decaying at large 
distances (greater than $d_N$) in any $N-$regular interval $A$ 
(it means an interval which does not contain any of the $I_{N,j}$
belonging to $S_N(Z,\,\lambda)$). 
It is proved by recursion over $n$ as follows:
\smallskip
\item{---}If $A\cap S_0=\emptyset$ it is obvious from the definition of $S_0$ 
that $G(\lambda,x,y)\le e^{-c|x-y|}$ if $|x-y|>d_0$.
\item{---}If the Green's function decays exponentially at distances larger
than $d_n$ in $n-$regular intervals and $A\cap S_{n+1}=\emptyset$,
then one can use
the resolvent identity (subscripts $I$ indicate Dirichlet boundary 
conditions at the endpoints of $I$): 
$$G_A(\lambda,x,y)\,=\,G_{I_{n,j}}(\lambda,x,y)+\sum_{z\in\partial I}\dot 
G_{I_{n,j}}(\lambda,x,z)G_A(\lambda,z,y)~,$$
where $\dot G$ is the derivative with respect to $z$. It 
allows us to express $G_A$ in terms of $G_{I_{n,j}}$ to which 
the recursion hypothesis applies, since the $I_{n,j}$ are of length $d_n$ 
and we want to prove exponential decay of $G(x,y)$ for $x,y$ distant
by more than $d_{n+1}$. 
\item{---}There is a similar bound for $\dot G$, because one can write 
$G$ as a function of $G_0$, the Green's function of the `free' operator 
$-\partial_x^2+W_\max$, using a resolvent identity:
$$G\,=\,G_0+(W_\max-W)G\,G_0~, $$
and $G_0$ as well as its derivatives behave like
$e^{-\sqrt{2-\alpha}|x-y|}$, for all $\lambda<M<2$.

Given the exponential decay of the Green's function, the behavior of the 
eigenfunctions follows from the formula:
$$e_j(x)\,=\,\sum_{z\in\partial A}\dot G_A(\lambda,x,z)e_j(z)~. $$
\QED

\REMARK We proved that the Green's function decays exponentially with
a certain rate $\kappa$. Hence this rate is the same for all
eigenfunctions $e_j$, $j\in\natural$.

%%%%%%%%%%%%%%%%%%%%%%%%%%%%%%%%%%%%%%%%%%%%%%%%%%%%%%%%%%%%%%%%%
\LIKEREMARK{Proof of \clm(projector)}We
use the following notations: $P(\cdot)$ is the spectral measure associated
with $L_Z$, $Z\in\Omega^*$, and $M>0$ is as in \clm(anderson). We show
that the restriction of $P_{N_\epsilon}$ to the subspace ${\rm
span}\{\tau_{z_j}\,,\,j\in\integer\}\backslash\{0\}$ has empty
kernel. Let
$\tau\in{\rm span}\{\tau_{z_j}\,,\,j\in\integer\}\backslash\{0\}$.
$$\eqalign{
M^2\|(1-P_{N_\epsilon})\tau\|_\Lambda^2\,&=\,M^2\int_M^\infty\langle
P(d\lambda)\tau,\,\tau\rangle_\Lambda
+M^2\sum_{j>N_\epsilon}\langle e_j,\tau\rangle^2_\Lambda\cr 
&\,\le\,\int_\real\lambda^2\langle
P(d\lambda)\tau,\,\tau\rangle_\Lambda+\OO(\epsilon )=\|L_Z\tau\|_\Lambda^2~,
} $$
hence $\|(1-P_{N_\epsilon})\tau\|_\Lambda^2\le B^2(Z)\|\tau\|_\Lambda^2$
where $B(Z)\equiv g_2(Z)/M$. By \clm(g1-g2bound), $B(Z)<1$ hence we
have $\inf\|P_{N_\epsilon}\tau\|_\Lambda\ge\bigl (1-B(Z)\bigr
)\|\tau\|_\Lambda\ge\beta \|\tau\|_\Lambda$ with $\beta >0$. This
proves that the map 
$$
P_{N_\epsilon}: {\rm span}\{\tau_{z_j},\,|j|\le D_\tau\}\to\bigoplus_{k\le
N_\epsilon}\HH_{\lambda_k}
$$
has trivial kernel. In addition, it is a map between
finite-dimensional spaces, and if $D_\tau>N_\epsilon$, then it is surjective.

 Now, define
$w=v-u_{Z(v)}$, $\tau\in{\rm
span}\{\tau_{z_j}\,,\,|j|\le D_\tau\}\backslash\{0\}$ such that
$P_{N_\epsilon}w=P_{N_\epsilon}\tau$. Recall that, by hypothesis, $\langle
w,\tau\rangle_\Lambda=0$, thus:
$$
\eqalign{
\|P_{N_\epsilon}w\|_\Lambda^2\,&=\,|\langle
w,\,P_{N_\epsilon}w\rangle_\Lambda|\cr
\,&=\,|\langle w,\,P_{N_\epsilon}\tau\rangle_\Lambda|\cr
\,&=\,|\langle w,\,(P_{N_\epsilon}-1)\tau\rangle_\Lambda|\cr
\,&\le\,\|w\|_\Lambda\|(P_{N_\epsilon}-1)\tau\|_\Lambda\cr
\,&\le\,{g_2(Z)\over
M}\|w\|_\Lambda\|\tau\|_\Lambda\,\le\,{g_2(Z)/M\over1-g_2(Z)/M}
\|w\|_\Lambda^2~,}
$$
using
$\|w\|_\Lambda\ge\|P_{N_\epsilon}w\|_\Lambda=\|P_{N_\epsilon}\tau\|_\Lambda
\ge\|\tau\|_\Lambda
+\|(1-P_{N_\epsilon})\tau\|_\Lambda\ge\|\tau\|_\Lambda(1-g_2(Z)/M)$.
The proof is complete, since we can use the bound of \clm(g1-g2bound).
\QED

%%%%%%%%%%%%%%%%%%%%%%%%%%%%%%%%%%%%%%%%%%%%%%%%%%%%%%%
\LIKEREMARK{Proof of \clm(positivity)}We let $Z=Z(v)$ and use the
following notations: $Z^*$ is
as in \clm(tube), $P(\cdot)$ is the spectral measure associated
with $L_{Z^*}$ and $\alpha>0$, $M>0$ are as in \clm(anderson). 
  
We start by proving the second inequality of Eq.\equ(positivity1). By
the Spectral Theorem, we have
$$
\eqalign{
\langle w,\,L_Z w\rangle_\Lambda\,&=\,
\langle w,\,L_{Z^*} w\rangle_\Lambda+\OO(\epsilon )\|w\|_\Lambda^2\cr
\,&=\,\int_\real\lambda\langle
P(d\lambda)w,\,w\rangle_\Lambda+\OO(\epsilon )\|w\|_\Lambda^2\cr
&\,\ge\,M\int_\real\langle
P(d\lambda)w,\,w\rangle_\Lambda+\int_{-\infty}^M\bigl
(\lambda-M\bigr )\langle P(d\lambda)w,\,w\rangle_\Lambda+\OO(\epsilon
)\|w\|_\Lambda^2\cr
&\,\ge\, M\|w\|_\Lambda^2+\sum_{j\le N_\epsilon}(\lambda -M)\langle
w,e_j\rangle_\Lambda^2
+\OO(\epsilon )\|w\|_\Lambda^2\cr
&\,\ge\,
M\|w\|_\Lambda^2+(-\alpha-M)\|P_{N_\epsilon}w\|_\Lambda^2+\OO(\epsilon
)\|w\|_\Lambda^2~.} 
$$ 
By \clm(projector), for $|Z|$ large, $\|P_{N_\epsilon}w\|_\Lambda\le
c_2(e^{-c_1|Z|})\|w\|_\Lambda$. Thus
$$
\langle w,\,L_Z w\rangle_\Lambda\,\ge\,\|w\|_\Lambda^2\Bigl
(M(1-\OO(e^{-c_1|Z|}))+\OO(\epsilon)\Bigr)\,\equiv\,M_1\|w\|_\Lambda^2~.
\EQ(mu1)$$
If $\Gamma=|Z|$ is sufficiently large and $\epsilon $ sufficiently
small, $M_1$ is positive.

To prove the first inequality of Eq.\equ(positivity1), we do similar
calculations: 
$$\eqalign{
\langle L_Zw,\,L_Zw\rangle_\Lambda\,&=\,\int_\real\lambda^2\langle
P(d\lambda)w,\,w\rangle_\Lambda+\OO(\epsilon )\|w\|^2_\Lambda\cr
\,&\ge\,M\int_\real\lambda\langle
P(d\lambda)w,\,w\rangle_\Lambda+\int_{-\infty}^M\bigl (\lambda -M\bigr
)\lambda \langle P(d\lambda)w,\,w\rangle_\Lambda\cr
\,&\,~~~~+\OO(\epsilon )\| w\|^2_\Lambda\cr
\,&\ge\, M\langle w,\,L_Z
w\rangle_\Lambda+(-\alpha-M)M\|P_{N_\epsilon}w\|_\Lambda^2+\OO(\epsilon
)\|w\|_\Lambda^2\cr
\,&\ge\, \Bigl(M(1-\OO(e^{-c_1|Z|})+\OO(\epsilon ))\Bigr )\langle
w,\,L_Zw\rangle_\Lambda ~,}$$ 
where we have used Eq.\equ(mu1).
If $|Z|=\Gamma $ is sufficiently large and $\epsilon $ sufficiently
small, the assertion follows.

We next prove Eq.\equ(positivity2). Since
$\partial_x(w^2)=2w\partial_xw\le(\partial_xw)^2+w^2$, supposing that
$\Lambda=[-K,K]$, we have:
$$
\eqalign{
\|\chi_\Lambda w\|_\infty^2\,&=\,\sup_{y\in\real}\int_{-\infty
}^y\!dx\,\partial_x(\chi_\Lambda^2w^2)\cr
\,&=\,\sup_{y\in\real}\int_{-\infty
}^y\!dx\,\bigl (2\chi_\Lambda
w^2\partial_x\chi_\Lambda+\chi_\Lambda\partial_x(w^2)\bigr )\cr
\,&\le\,\sup_{y\in\real}\int_{-\infty
}^y\!dx\,\chi_\Lambda\bigl
((2\partial_x\chi_\Lambda+1)w^2+(\partial_xw^2)\bigr )\cr
\,&\le\,C_1\sup_{y\in\real}\int_{-\infty }^y\!d\mu\,\bigl
(2w^2+(\partial_xw^2)\bigr )\cr
\,&\le\,C_1\int_\real\!d\mu\,\bigl (2w^2+(\partial_xw)^2\bigr )~.
}
$$
It only remains to prove that for some $C_2>0$,
$I\equiv\int_\real\!d\mu\,wL_Zw-C_2\int_\real\!d\mu\,\bigl((\partial_xw)^2+w^2
\bigr )>0$. We have:
$$
\eqalign{
I\,&=\,\bigl(1-C_2)\int_\real\!d\mu\,wL_Zw
+C_2\int_\real\!d\mu\,\bigl (wL_Zw-(\partial_xw)^2-2w^2\bigr )\cr
\,&=\,\bigl(1-C_2)\int_\real\!d\mu\,wL_Zw
-C_2\int_\real\!d\mu\,\bigl (2+V''(u_Z)\bigr )w^2\cr
\,&\ge\,(1-C_2)M_1\|w\|_\Lambda^2-C_2\sup_{|x|\le
1}|2+V''(x)|\|w\|_\Lambda^2\cr
\,&\ge\,\bigl (M_1-(M^*+M_1)C_2\bigr )\|w\|_\Lambda^2~,
}
$$ 
where we have defined $M^*\equiv\sup_{|x|\le
1}|2+V''(x)|$. The proof is complete if one chooses $C_2<M_1/(K+M_1)$.
\QED

%%%%%%%%%%%%%%%%%%%%%%%%%%%%%%%%%%%%%%%%%%%%%%
\SECT(geometry)Proofs of the geometric structure
%%%%%%%%%%%%%%%%%%%%%%%%%%%%%%%%%%%%%%%%%%%%%

We first prove the existence of an orthogonal
coordinate system adapted to the problem.
%%%%%%%%%%%%%%%%%%%%%%%%%%%%%%%%%%%%%%%%%%%%%
\LIKEREMARK{Proof of \clm(tube)}We fix $z_j$, $|j|>D_\tau+1$ such that
$\|v-u_Z\|_\Lambda<\infty$ and we apply
the Implicit Function Theorem to the function  
$$
\Bigl(\FF(v,z_{-D_\tau-1},\dots,z_{D_\tau+1})\Bigr)_j\,=\,\langle v-u_Z,\,
\tau_{z_j}\rangle~,\;{\rm for~}|j|\le D_\tau+1~.
$$
We can check that the hypotheses are satisfied:
\item{1)}$\FF(u_Z,Z)=0~,$
\item{2)}$({\rm D}\FF(u_Z,Z))=-\SS$
where $\SS$ is as in \clm(ss) (with $N=D_\tau+1$), hence it has
bounded inverse.

To prove the second part of the claim, we note that $\Omega^*$ has
measure one, thus is it dense in $\Omega_\Gamma$ and, for all $\epsilon'>0$,
for each $Z\in\Omega_\Gamma$, the set $\{\tilde
Z\in\Omega_\Gamma:\,|\tilde Z-Z|<\epsilon'\}$
is an open subset of $\Omega_\Gamma$ and thus contains an
$Z^*\in\Omega^*$. By continuity, there is a $C$ such that
$\|(L_Z-L_{Z^*})w\|_\Lambda/\|w\|_\Lambda<C\epsilon'=\epsilon$.
\QED

Before proceeding to the proofs of \clm(cone) we put Eq.\equ(motion2)
in a more compact form. We have, using Eq.\equ(lineq):
$$
\eqalign{
\partial_t Z(v)\,&=\,\SS^{-1}\langle\LL (v),\,\tau_Z\rangle_\Lambda\cr
\,&=\, \SS^{-1}\langle\LL (w+u_Z),\,\tau_Z\rangle_\Lambda\cr
\,&=\, \SS^{-1}\langle\LL (u_Z)-L_Z w +w^2 r(w,u_Z),\,\tau_Z\rangle_\Lambda\cr
\,&=\, \SS_1^{-1}\langle\LL (u_Z),\tau_Z\rangle_\Lambda
+(\SS^{-1}-\SS_1^{-1})\langle\LL (u_Z),\tau_Z\rangle_\Lambda\cr
&~~~~+\SS^{-1}\langle -L_Z w +w^2 r(w,u_Z),\,\tau_Z\rangle_\Lambda\cr
\,&\equiv\, \PP^{(1)} + \PP^{(2)} + \PP^{(3)}~.\cr
}
\EQ(decompose1)
$$
The equation for $w$ takes the form
$$
\eqalign{
\partial_t w\,&=\,\LL (u_Z)-L_Z w +w^2 r(w,u_Z)-\DZ u_Z \cdot \partial_t
Z(w+u_Z)\cr
\,&=\,\LL (u_Z)-\DZ u_Z\cdot \bigl (\SS_1^{-1} \langle \DZ u_Z ,\LL
(u_Z)\rangle_\Lambda \bigr )\cr 
&~~~~-L_Z w +w^2 r(w,u_Z)\cr
&~~~~ +\DZ u_Z \cdot \bigl ( -\partial_t
Z(w+u_Z) + \SS_1^{-1} \langle \DZ u_Z ,\LL
(u_Z)\rangle_\Lambda \bigr )\cr
\,&\equiv\, \QQ -L_Z w +w^2 r(w,u_Z) +\DZ u_Z\cdot \bigl
(-\PP^{(2)}-\PP^{(3)}\bigr )~.
}\EQ(decompose2)
$$

\CLAIM Lemma(sbound)
There exist constants $c_1>0$, $c_2>0$, $c_3$, and $c_4>0$ such that for
sufficiently small $\tub$, the following holds:
$$
|\SS_1-E\cdot 1|\,\le\, c_2(\|w\|_\Lambda+e^{-c_1|Z|}) ~,\qquad
|\SS^{-1}-\SS_1^{-1}|\,\le\, c_3\|w\|_\Lambda~, 
$$ 
where $E=c_4\int_{-1}^1 dy\,\sqrt{2V(y)}=c_4\int_{-\infty}^\infty
dx\,\bigl(\psi'(x)\bigr)^2$.

\PROOF The first claim is proved with the following argument: The
off-diagonal elements of $\SS_1$ are
of order $\|w\|_\Lambda$, and the diagonal ones satisfy the bound:
$$\eqalign{
|(\SS_1)_{ii}|&\,\le\,c_4\int_{c_i^++1}^{c_{i+1}^+-1}\!dx\,|\partial_{z_i}u_Z
\partial_xu_Z|\cr
\,&\le\,c_2e^{-c_1|Z|}+
c_4\int_{c_i^++1}^{c_{i+1}^+-1}\!dx\,|\partial_xu_Z|^2\cr
\,&=\,c_2e^{-c_1|Z|}+c_4\int_{c_i^++1}^{c_{i+1}^+-1}\!dx\,\sqrt{2V(u_Z)}\cr
&\,\le\,c_2e^{-c_1|Z|}+E~,
} $$
using Lemma 7.8 of [CP1] to compare $\partial_{z_j}u_Z$ with
$\partial_xu_Z$.

The second claim comes from the following estimation (defining
$\SS_2=\SS_1-\SS$): 
$$
\eqalign{
\SS^{-1}-\SS_1^{-1}\,&=\,\SS^{-1}\bigl (1-(\SS_1-\SS_2)\SS_1^{-1}\bigr )\cr
\,&=\,\SS^{-1}\SS_2\SS_1^{-1}\,\le\,C\|w\|_\Lambda~,
}
$$
because of \clm(ss) and using that $\left (\SS_2\right )_{i,j}=\langle
w, \partial_{z_j}\tau_{z_i}\rangle_\Lambda\le C\|w\|_\Lambda$.
\QED

\clm(sbound) is used to prove estimates on expressions appearing in
Eq.\equ(decompose1) and Eq.\equ(decompose2):

\CLAIM Lemma(ppbound)
There exists constants $c_1>0$, $c_2>0$, and $c_3>0$ such that for
sufficiently small $\tub$, the following holds:
$$
\eqalign{
\sup_{|j|\le D_\tau}|\PP^{(1)}_j|\,&\le\, c_1g_1(Z)~,\cr
\sup_{|j|\le D_\tau}|\PP^{(2)}_j+\PP^{(3)}_j|\,&\le\, c_2\|w\|_\Lambda\bigl
(g_1(Z)+g_2(Z)+\|\chi_\Lambda w\|_\infty +\epsilon\bigr )~.
}
$$

\PROOF The bound on $\PP^{(1)}$ is obvious from its definition, from
the definition of $g_1(Z)$ and from \clm(sbound).

The bound on $\PP^{(2)}+\PP^{(3)}$ follows from (see Eq.\equ(infty-bound)):
$$
|\langle L_Zw,\tau_{z_j}\rangle_\Lambda|\,=\,
|\langle w,L_Z\tau_{z_j}\rangle_\Lambda|\,\le\,C\|w\|_\Lambda g_2(Z)~,
$$
$$
|\langle w^2r(w,u_Z),\tau_{z_j}\rangle_\Lambda|\,\le\,\|r\|_\infty
\|\tau_{z_j}\|_\infty \|w\|^2_\Lambda\,\le\,C(\|\chi_\Lambda w\|_\infty
+\epsilon)\|w\|_\Lambda~,
$$
and from \clm(sbound).
\QED

We are now prepared to give the proof of \clm(cone):

%%%%%%%%%%%%%%%%%%%%%%%%%%%%%%%%%%%%%%%%%%%%%
\LIKEREMARK{Proof of \clm(cone)}We start by expanding the first term
of the l.h.s.\ of Eq.\equ(cone-ineq). Denoting $Z=Z(v)$ and $w=v-u_{Z(v)}$
and using Eq.\equ(motion2) and Eq.\equ(lineq) as well as
\clm(positivity), we get
$$
\eqalign{
\HALF\partial_t\|w\|_Z^2\,&=\,\langle \dot
w,L_Zw\rangle_\Lambda-\int_\real\!d\mu\,w^2V''(u_Z)\left 
(\DZ u_Z\cdot\dot Z\right ) \cr
\,&=\, \langle\QQ,L_Zw\rangle_\Lambda -\|L_Zw\|_\Lambda +\langle w^2
r(w,u_Z),L_Zw\rangle_\Lambda\cr
\,&\,~~~~+\langle
L_Zw,D_Zu_Z\cdot(-\PP^{(2)}-\PP^{(3)})\rangle_\Lambda\cr
\,&\,~~~~ -\int_\real\!d\mu\,w^2V''(u_Z)\left
(\DZ u_Z\cdot \bigl(\PP^{(1)}+\PP^{(2)}+\PP^{(3)}\bigr )\right )\cr
\,&\le\,{\textstyle{1\over
4}}\|L_Zw\|_\Lambda^2+\|\QQ\|_\Lambda^2-\|L_Zw\|_\Lambda\cr
\,&\,~~~~+C\bigl (\|\chi_\Lambda
w\|_\infty\|w\|_\Lambda\|L_Zw\|_\Lambda
+\|(1-\chi_\Lambda)w\|_\Lambda\|w\|_\Lambda\|L_Zw\|_\Lambda\bigr )\cr
\,&\,~~~~+C\bigl
(\|L_Zw\|_\Lambda+\|w\|_\infty \|w\|_\Lambda\bigr )\|w\|_\Lambda\bigl
(g_1(Z)+g_2(Z)+\|\chi_\Lambda w\|_\infty+\epsilon  \bigr )\cr
\,&\,~~~~+C\|w\|_\Lambda^2 g_1(Z)\cr
\,&\le\,\|L_Zw\|^2\bigl (-{\textstyle{3\over
4}}+C(\|\chi_\Lambda w\|_\infty+\epsilon ) +C(1+\|w\|_\infty
)\cr
\,&\,~~~~\times(g_1(Z)+g_2(Z)+\|\chi_\Lambda w\|_\infty+\epsilon  )\bigr )
+\|\QQ\|_\infty^2~,}
$$
where we have used \clm(positivity) and \clm(ppbound). Taking $\tub$
and $\epsilon $ sufficiently small, we get 
$$
\HALF\partial_t\|w\|_Z^2\,\le\,-\HALF\|L_Zw\|^2_\Lambda+\|\QQ\|_\infty
^2~.\EQ(bound-dt-w)
$$

We next expand and estimate the time derivative of $g_1(Z)$
$$
\eqalign{
\HALF\partial_t g_1^2(Z)\,&=\,\sum_{-D_\tau\le j,k\le
D_\tau}\langle\LL(u_Z),\tau_{z_j}\rangle_\Lambda\bigl
(\langle
L_Z\partial_{z_k}u_Z,\tau_{z_j}\rangle_\Lambda
\cr
\,&\,~~~~+\langle\LL(u_Z),\partial_{z_k}\tau_{z_j}\rangle_\Lambda\bigr
)\bigl (\PP^{(1)}_k + \PP^{(2)}_k + \PP^{(3)}_k\bigr )~.}$$
We have:
$$
\eqalign{
|\langle L_Z\partial_{z_k}u_Z,\tau_{z_j}\rangle_\Lambda|\,&\le\,
|\langle\partial_{z_k}u_Z,L_Z\tau_{z_j}\rangle_\Lambda|\,\le\,Cg_2(Z)~,\cr
|\langle\LL(u_Z),\tau_{z_j}\rangle_\Lambda|\,&\le\,
\|\LL(u_Z)\|_\Lambda\|\partial_{z_k}\tau_{z_j}\|_\Lambda\,\le\,Cg_1(Z)~.
}
$$ 
Hence, using \clm(ppbound), we get 
$$
\HALF\partial_t g_1^2(Z)\,\le\,Cg_1(Z)\bigl (g_1(Z)+g_2(Z)\bigr )\bigl
(g_1(Z)+\|L_Zw\|_\Lambda(g_1(Z)+g_2(Z)+\|\chi_\Lambda w\|_\infty
+\epsilon)\bigr )~.\EQ(bound-dt-g)
$$

Summing Eq.\equ(bound-dt-w) and Eq.\equ(bound-dt-g) and using
$\|\QQ\|_\infty \le Cg_1(Z)$, we have:
$$
\eqalign{
\HALF\partial_t\bigl (\|w\|_Z^2-Bg_1^2(Z)\bigr
)\,&\le\,-\HALF\|L_Zw\|_\Lambda^2+Cg_1^2(Z)+CBg_1(Z)\bigl
(g_1(Z)+g_2(Z)\bigr )\cr
\,&\,~~~~
\times\bigl(g_1(Z)+\|L_Zw\|_\Lambda(g_1(Z)+g_2(Z)+\|\chi_\Lambda
w\|_\infty+\epsilon )\bigr)\cr
\,&\le\,-{\textstyle{1\over 4}}\bigl
(\|L_Zw\|_\Lambda^2-M_2Bg_1^2(Z)\bigr )\cr 
\,&\,~~~~-{M_2\over 4}Bg_1^2(Z)
+Cg_1^2(Z)\bigl (1+B(g_1(Z)+g_2(Z))\bigr )\cr
\,&\,~~~~+CB^2g_1^2(Z)\bigl (g_1(Z)+g_2(Z)\bigr )^2\bigl
(g_1(Z)+g_2(Z)+\|\chi_\Lambda w\|_\infty+\epsilon \bigr )^2\cr
\,&\le\,-{\textstyle{1\over 4}}\bigl
(\|L_Zw\|_\Lambda^2-M_2Bg_1^2(Z)\bigr )\cr 
\,&\,~~~~+Cg_1^2(Z)\FF(B)
~,}
$$
where 
$$
\eqalign{
\FF(B)\,&=\,\bigl (\AA_1B^2+\AA_2B+1\bigr )~,\cr
\AA_1\,&=\, (g_1(Z)+g_2(Z)\bigr )^2\bigl
(g_1(Z)+g_2(Z)+\|\chi_\Lambda w\|_\infty+\epsilon \bigr )^2~,\cr
\AA_2\,&=\,(g_1(Z)+g_2(Z))-{M_2\over 4C}~.
}
$$
We take $\tub$ and $\epsilon $ so small that
$g_1(Z)+g_2+\|\chi_\Lambda w\|_\infty+\epsilon 
<{\textstyle{1\over 4}}$ and $g_1(Z)+g_2(Z)<M_2/8C$. Thus, 
$\FF(B)\le1-M_2B/(8C)+(M_2/(8C))^2(1/4)^2B^2$, hence it is
negative for $1-\sqrt{3}/2<M_2B/(64C)<1+\sqrt{3}/2$. 
\QED

We finally give proof of \clm(speed). Note that
in \clm(speed), we assume that
$v=w+u_{Z(v)}\in\ZZ$ hence $\|\chi_\Lambda w\|_\infty \le C\|w\|_Z\le
Cg_1(Z)$ which is smaller than $\sigma$.

\LIKEREMARK{Proof of \clm(speed)}We start by recalling
Eq.\equ(decompose1), with the shorthand $Z=Z(v)$ and with the convention
that repeated indices are summed over:
$$
\eqalign{
\partial_tZ(v)\,&=\,\bigl (\SS_1^{-1}\bigr
)_{j,i}\langle\LL(u_Z),\,\tau_{z_i}\rangle+\bigl
(\SS^{-1}-\SS_1^{-1}\bigr )_{j,i}\langle\LL(u_Z),\,\tau_{z_i}\rangle\cr
\,&\,~~~~+\bigl (\SS^{-1}\bigr )_{j,i}\langle
-LZw,\tau_{z_i}\rangle+\bigl (\SS^{-1}\bigr )_{j,i}\langle
w^2r(w,u_Z),\tau_{z_i}\rangle 
~.
}\EQ(motion3)
$$
Using \clm(sbound), we find 
$$
|(\SS_1)_{ii}|\,\le\,E+Ce^{-c_1|Z|}~.
$$
It follows that the first term of Eq.\equ(motion3) is equal to
$E\langle\LL(u_Z),\tau_{z_j}\rangle+\OO(e^{-c_1|Z|})g_1(Z)$.
The second term is also estimated using \clm(sbound)
$$
\eqalign{
\SS^{-1}-\SS_1^{-1}\,&\le\,C\|w\|_2\,\le\,Cg_1(Z)~,
}
$$
thus, the second term in Eq.\equ(motion3) is equal to
$\OO(e^{-c_1|Z|})\OO(g_1(Z))$. The third term is estimated as in the
proof of \clm(ppbound) replacing $\sigma$ by $g_1(Z)$ everywhere. This
completes the proof of \clm(speed).
\QED

%%%%%%%%%%%%%%%%%%%%%%%%%%%%%%%%%%%%%%%%%%%%%%%%%%%%%%%%%%%%%%%%%%%%%%%%
\SECT(universal-proof)Proof of \clm(universal)
%%%%%%%%%%%%%%%%%%%%%%%%%%%%%%%%%%%%%%%%%%%%%%%%%%%%%%%%%%%%%%%%%%%%%%%%

First, we study the simpler case of the collapse of a function with
two kinks separated by a distance $\Gamma$ and then, we compare with the
evolution of the many-kink solution.

\CLAIM Lemma(two-kinks)
For sufficiently large $\Gamma>P_0$ and $\Gamma_0>\Gamma$, 
there are constants $\kappa_0>0$ and $\epsilon_0 >0$
such that the following is true.
If $v_0=u^*_\Gamma$ where $u^*_\Gamma$ is given by Eq.\equ(kink) with
$z_0=-\infty$, $z_1=0$, $z_2=\Gamma$, and  $z_3=\infty $ with the
convention that $\phi_P=\psi$ if $P=\infty $. Let $v_t$ be the corresponding
solution of Eq.\equ(gl). Then there is a $T_p<\infty$ such that
$v_{T_p}(x)>\epsilon_0>0$ for all $x\in\real$.

\PROOF We use the parabolic maximum principle
together with the existence of moving front solutions for the
Eq.\equ(gl).  
A front is a function $f_s(x-st)$, where $s$ is a fixed
number and $f_s(x)$ solves:
$$\partial_x^2f_s+s\partial_xf_s+V'(f_s)\,=\,0~.\EQ(fric)$$
In the mechanical interpretation shown
in \fig(images/classical.ps), $f_s(x)$ is an 
oscillating trajectory subject to a constant friction $s$.

Let $\GR>\Gamma$, let $\phi_{\GR}=\phi_P$ with $P=\GR$, and let $f_s$
be the solution of Eq.\equ(fric) with
initial values $f_s(\GR)=\phi_{\GR}(\GR)=0$ and
$f_s'(\GR)=\phi_{\GR}'(0)<0$ satisfies:
\item{1)}$f_s(x)<\phi_{\GR}(x)$ for $0<x<\GR$ and for all $s>0$.
\item{2)}For sufficiently small $s>0$, there exist $\GL\ne\GC$ with
$\GL<-\GR$  and $\GC<0$ such that $f_s(\GL)=f_s(\GC)=0$.

The first claim is a consequence of the following argument: Integrating
Eq.\equ(fric) from the initial values at $x=\GR$, we see that
in a small neighborhood of $\GR$, we have 
$f_s(x)<\phi_{\GR}(x)$. Let $A$ be given by the equation $\GR=P(A)$, {\it
cf.~}\clm(persol), and let $x_0$ such that $\phi_{\GR}(x_0)=-A$.
Then, for $x_0<x<\GR$,
$$\eqalign{\phi_{\GR}(x)'\,&=\,\sqrt{2}\sqrt{V(A)-V(\phi_{\GR}(x))}~,\cr
f_s(x)'\,&>\,\sqrt{2}\sqrt{V(A)-V(f_s(x))}~.}\EQ(slope)$$
If we suppose that there exists an $x^*$, $x_0<x^*<\GR$
such that $f_s(x^*)=\phi_{\GR}(x^*)$, since $f_s(x)<\phi_{\GR}(x)$
near $x=\GR$,
we see that $f_s'(x^*)<\phi_{\GR}'(x^*)$ which is a contradiction with
Eq.\equ(slope). In the interval $(0,x_0)$ the same argument applies
with opposite signs for the square roots in Eq.\equ(slope). Hence
$f_s$ does not intersect $\phi_{\GR}$ {\it i.e.,} $f_s$ lies below
$\phi_{\GR}$ in the interval $(0,\GR)$.

The second claim follows from the observation that
$f_s(x)\to\phi_{\GR}(x)$ when $s\to 0$ hence, by continuity, there
exist such zeros of $f_s$ for small $s$.

Furthermore, we have that
$u_{\Gamma^*}(x)\ge\psi_{\rm R}(x)\equiv\psi(-(x-\Gamma))$ and 
$u_{\Gamma^*}(x)\ge\psi_{\rm C}(x)\equiv\psi(x-\GC)$.
Hence, by the maximum principle,
$$h(x,T+t)\,>\,{\rm max}\bigl (f_s(x-st),
\psi_{\rm L}(x),
\psi_{\rm C}(x)\bigr )~,
$$
for all $t>0$, and for $f_s$ satisfying 1) and 2). In particular, for
$T_p=(\GR-\GC)/s$, the function
$h(x,T+T_p)$
is strictly positive (see \fig(images/newfront2.ps),
\fig(images/newfront3.ps), and
[CE], p.149, Example 4).
\QED
%\figurewithtexplus images/newfront2.ps images/newfront2.tex 8.1 12.0 -1.2
%The moving front $f_s$, the stationary solutions $\psi_{\rm R}(x)$,
%$\psi_{\rm C}(x)$, and $u_{\Gamma^*}$.\cr
%\figurewithtexplus images/newfront3.ps images/newfront3.tex 8.1 12.0 -1.2 The
%position of the front $f_s$ at time $T_p=(\GR-\GC)/s$,
%relative to $\psi_{\rm R,C}(x)$.\cr

\LIKEREMARK{Proof of \clm(universal)}We first remark that $v_0$
restricted to the interval
$I_1\equiv [z_j-\Gamma_0/3,z_{j+1}+\Gamma_0/3]$ is close
to the two-kinks function
$u_{\tilde Z}$, with $\tilde Z=\{z_j,z_j+\Gamma\}$. 
The evolution of
$\tilde u_{\tilde Z}$ has been described in \clm(two-kinks), and is known to
lead to a collapse.
We now show that the evolutions of $\tilde u_{\tilde Z}$ and $u_Z$ remain
close to each other for a time longer than the time $T_p$ needed for collapse.

To perform the comparison, we consider the functions 
$f_0=v_T=w+u_Z$ and $\tilde f_0=\tilde w+u_{\tilde Z}$. 
By Duhamel's Principle:
$$\eqalign{
f_t(x)-\tilde f_t(x)\,&\equiv\,f(x,t)-\tilde f(x,t)\cr
\,&=\,\bigl
(e^{t\partial_x^2}(f_0-\tilde f_0)\bigr )(x)+\int_0^t
ds\,\Bigl (e^{(t-s)\partial_x^2}\bigl (V'(f_s)-V'(\tilde f_s)\bigr
)\Bigr )(x)~.}\EQ(Duhamel)$$  
Let $I_2\equiv [z_j,z_j+\Gamma]\subset I_1$. Let $\mu$ be an
absolutely continuous measure on $\real$ such
that $\mu(I_2)=1-\epsilon_0$ and $\mu(\real\backslash I_2)=\epsilon_0$
with $\epsilon _0>0$ and let $\|\cdot\|_r$ be the
$L^r(\real,d\mu)-$norm. 
If $\Delta _t$ denotes the l.h.s\ of Eq.\equ(Duhamel), we have
that $\Delta _0(x)=0$ if $x\in I_1$ and $|V'(f_s)-V'(\tilde
f_s)|\le\kappa|\Delta_s|$ for some $\kappa>0$ because $V'$ is in ${\cal
C}^1$. Using that
$e^{t\partial_x^2}$ is an $L^p-$contractive semi-group (see [RS2], p.255),
we have that 
$$\eqalign{
\|\Delta_t\|_r \,&\le\,\|e^{t\partial_x^2}\Delta _0\|_r+\kappa\int_0^t
ds\,\|e^{(t-s)\partial_x^2}\Delta_s\|_r\cr
\,&\le\,\|\Delta _0\|_r+\kappa\int_0^t
ds\,\|\Delta_s\|_r\cr
\,&\le\,\delta (\Gamma_0)+\kappa\int_0^t ds\,\|\Delta_s\|_r\cr
\,&\le\,\delta (\Gamma_0)e^{\kappa t}~,}
$$
for all $p\ge 1$, $t<T_p$, and with $\delta (\Gamma_0)\to 0$ when
$\Gamma_0\to\infty$. The last line follows from Gronwall's
Lemma. Thus, for $\Gamma_0$ sufficiently large, $\|\delta
_{T_p}\|_r<\epsilon/2$ for all $r\ge 1$. Let ${\bf 1}_{I_2}$ be the
indicator function of the interval $I_2$, then by H\"older's Inequality:
$$
\eqalign{
\Bigl (\int_{I_2}d\mu\,|\Delta _{T_p}|^r\Bigr )^{1/r}\,&=\,\Bigl
(\int_\real d\mu\,{\bf 1}_{I_2}|\Delta_{T_p}|^r\Bigr )^{1/r}\cr
\,&\le\,\Bigl (\int_\real d\mu\,{\bf 1}_{I_2}\Bigr )^{1/q}\Bigl (\int_\real
d\mu\, |\Delta_{T_p}|^s \Bigr )^{1/s}\cr
\,&<\,(1-\epsilon_0)^{1/q}\epsilon /2~,
}
$$
if $1\le r<q$. Thus, since $\|f\|_r\to\|f\|_\infty $ when $r\to
\infty $ (see [R] p.71), we get that $\sup_{x\in I_2}|\Delta
_{T_p}|<\epsilon/2$, and, since $|f^*_{T_p}(x)|>\epsilon$ by
\clm(universal), we find 
$|f_{T_p}(x)|>\epsilon/2$ for $x\in I_2$.
\QED

%%%%%%%%%%%%%%%%%%%%%%%%%%%%%%%%%%%%%%%%%%%%%%
\SECT(periodic.proof)Proof of \clm(persol)
%%%%%%%%%%%%%%%%%%%%%%%%%%%%%%%%%%%%%%%%%%%%%

Instead of the convention settled in \clm(persol), we shall choose
the more symmetric definition: $\phi_P(x)$ has a minimum at
$x=0$, {\it i.e.}, $\phi_P(0)=-A$.
We seek particular solutions of the equation $\LL(u)=0$. In the
mechanical interpretation of a free particle moving in the potential
$V$ without friction, $u(x)$ is the position of the particle at time
$x$ (see \fig(images/classical.ps)). Intuitively, it is clear that if
the particle starts at rest from a position $u(0)$, with
$-1<u(0)=-z<0$, its trajectory will oscillate around $0$ with a
certain period $2P$. Looking for a relation between $P$ and $A$,
we show that if $u$ solves the initial value problem
$$\eqalign{&u''\,=\,-V'(u) ~,\cr
&u(0)\,=\,-A~,\;u'(0)=0~,}\EQ(classequ)$$ 
then there exists a (minimal) $P(A)$ such that $u(P(A)/2)=0$. We can
transform the equation \equ(classequ) into:
$\HALF((u')^2)'=-(V(u))'$ (supposing $u'\not\equiv 0$), which, after
integration, becomes $(u')^2=-2(V(u)-V(-A))$, where the integration
constant was set to $V(-A)$ in order to match the condition
$u'(0)=0.$ When $-A\le u<0$, $V(-A)\ge V(u)$, hence we can take the
square root:  
$$u'\,=\,\sqrt{2}\sqrt{V(-A)-V(u)}~.$$ 
The r.h.s.\ is invertible if $-A<u\le 0,$ yielding an equation for
the inverse function $x(u)$:
$$
x'(u)={1\over\sqrt{2}\sqrt{V(-A)-V(u)}}~.
$$ 
There will be a solution
satisfying the boundary condition $u(0)=-A$ and the periodicity
condition $u(P(A)/2)=0$ if and only if the integral 
$${1\over\sqrt{2}}\int_{-A}^0
{ds\over\sqrt{V(-A)-V(s)}}\,=\,x(0)-x(-A)\,=\,P(A)/2-0\EQ(amplitude-integral)
$$ 
exists. This is an elliptic integral of the first kind, which 
is an analytic bijection from $(0,1)$ onto $(P_0,\infty)$
(see \fig(images/elliptic.ps) and, {\it e.g.}, [A], p.322-324).
%\figurewithtexplus images/elliptic.ps images/elliptic.tex 8.5 10.0 -0.8
%The period $2P$ as a function of the amplitude $A$.\cr

We have described the solution $u$ of Eq.\equ(classequ) on the
interval $[0,P(A)/2]$, and we can 
indeed check that it extends to a periodic function. The equation 
$\HALF (u')^2+V(u)=V(-A)=V(A)$ together with $u''=-A'(u)$ leads to 
the existence of a number $P^*(A)$ where 
$$u\bigl(P^*(A)/2\bigr)\,=\,A~,\;u'\bigl(P^*(A)/2\bigr)\,=\,0~,\;u''\bigl(
P^*(A)/2\bigr)\,<\,0~,$$ 
{\it i.e.}, a maximum of $u$ of height $A$.
This number $P^*(A)$ is determined by (recall that $V$ is even):  
$$\eqalign{&x(z)-x(0)\,=\,P^*(A)/2-P(A)/2\,=\,{1\over\sqrt{2}}\int_0^{z} 
{ds\over\sqrt{V(A)-V(s)}}\cr
\,&=\,{1\over\sqrt{2}}\int_z^0 {ds\over\sqrt{V(A)-V(s)}}\,=\,P(A)/2~.}$$
The initial value problem \equ(classequ)
is invariant under $x\rightarrow -x$, 
hence the solution is even. Altogether, we found the behavior of 
$u$ on the interval $[-P(A),P(A)]$ and since
$u\bigl(-P(A)\bigr)=u\bigl(P(A)\bigr)=A$ 
and $u'\bigl(-P(A)\bigr)=u'\bigl(P(A)\bigr)=0$, periodic copies will match 
together.
\QED

\REMARK An additional property of the integral \equ(amplitude-integral)
that is useful for our analysis is the following: For sufficiently
large $P$, there exist constants $c_1>0$ and $c_2>0$ such that
$$
1-A(P)\,\le\,c_2e^{-c_1P}~.\EQ(exponent-amplitude)
$$

%%%%%%%%%%%%%%%%%%%%%%%%%%%%%%%%%%%%%%%%%%%%%%%%%
%%%%%%%%%%%%%%%%%%%%%%%%%%%%%%%%%%%%%%%%%%%%%%%%%

\SECTIONNONR References
 
\eightpoint
\raggedright
\widestlabel{[XXX]}
\vskip 15pt

\ref
  \no A
  \by Arfken, G.
  \book Mathematical Methods for Physicists, 3d edition
  \publisher Orlando, Academic Press
  \yr 1985
\endref

\ref
\no BDG
  \by Bray, A.J., Derrida, B. and C. Godr\`eche
  \paper Non-trivial Algebraic Decay in a Soluble Model of Coarsening
  \jour Europhys. Lett.
  \vol 27 {\rm (3)}
  \pages 175-180
  \yr 1994
\endref

\ref
%\no BD2
%  \by Bray, A.J., Derrida, B. and C. Godr\`eche
%  \paper Non-trivial Exponents in the Zero Temperature Dynamics of the
%  1D Ising and Potts Models
%  \jour J. Phys. A: Math. Gen. 
%  \vol 27
%  \pages L357-L361
%  \yr 1994
%\endref

\ref
\no BD
  \by Bray, A.J. and B. Derrida
  \paper Exact Exponent $\lambda$ of the Autocorrelation Function for 
a Soluble Model of Coarsening
  \jour Phys. Rev. E 
  \vol 51 {\rm (3)}
  \pages 1633-1636
  \yr 1995
\endref

\ref
  \no CE
  \by Collet, P. and J.-P. Eckmann
  \book Instabilities and Fronts in Extended Systems
  \publisher Princeton, Princeton University Press
  \yr 1990
\endref

\ref 
\no CP1
  \by Carr, J. and R.L. Pego
  \paper Metastable Patterns in Solutions of 
$u_t=\epsilon^2u_{xx}-f(u)$
  \jour Comm. Pure. Appl. Math.
  \vol XLII
  \pages 523-576
  \yr 1989
\endref

\ref 
\no CP2
  \by Carr, J. and R.L. Pego
  \paper Invariant Manifolds for Metastable Patterns in 
$u_t=\epsilon^2u_{xx}-f(u)$
  \jour Proc. Roy. Soc. Edinburgh
  \vol 116A
  \pages 133-160
  \yr 1990
\endref

%\ref
%  \no F
%  \by Friedman, A.
%  \book Partial Differential Equations of Parabolic Type
%  \publisher Englewood Cliffs, Prentice-Hall
%  \yr 1964
%\endref

\ref 
\no FSW
  \by Fr\"ohlich, J., Spencer, T. and P. Wittwer
  \paper Localization for a Class of One Dimensional Quasi-Periodic\HB 
Schr\"odinger Operators
  \jour Comm. Math. Phys.
  \vol 132
  \pages 5-25
  \yr 1990
\endref

\ref 
\no G
  \by Gallay, Th.
  \book Existence et stabilit\'e des fronts dans l'\'equation de
  Ginzburg-Landau \`a une dimension
  \publisher PhD Thesis, University of Geneva
  \yr 1994
\endref

\ref
  \no R
  \by Rudin,W.
  \book Real and Complex Analysis
  \publisher Singapore, McGraw-Hill
  \yr 1987
\endref

\ref
  \no RS2
  \by Reed, M. and B. Simon
  \book Methods of Modern Mathematical Physics, II: Fourier Analysis,
  Self-Adjointness
  \publisher San Diego, Academic Press
  \yr 1975
\endref

\ref
  \no RS4
  \by Reed, M. and B. Simon
  \book Methods of Modern Mathematical Physics, IV: Analysis of Operators  
  \publisher San Diego, Academic Press
  \yr 1978
\endref

\ref
  \no S
  \by Spencer, T.
  \paper The Schr\"odinger Equation with a Random Potential
  \inbook Critical Phenomena, Random Fields, Gauge Theories. Les Houches 1984
  \publisher Amsterdam, North Holland
  \yr 1986
\endref

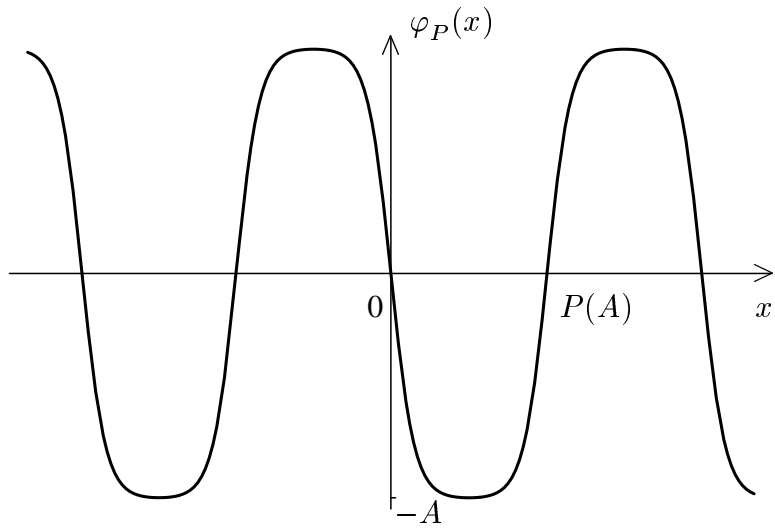
\bye